\documentclass[]{iopart}

\usepackage{iopams}  
\usepackage{subfigure}
\usepackage{amssymb}
\usepackage[normalem]{ulem}
\usepackage{color}
\usepackage[usenames,dvipsnames]{xcolor}
\usepackage{graphics}
\usepackage{graphicx}
\newcommand{\frep}{f_{rep}}
\newcommand{\fceo}{f_{ceo}}

\begin{document}

\review[High repetition rate fsEC based XUV frequency combs]{XUV Frequency Combs via Femtosecond Enhancement Cavities}

\author{Arthur K Mills, TJ Hammond, Matthew HC Lam and David J Jones}

\address{Department of Physics and Astronomy, University of British Columbia,
         \\ Vancouver, British Columbia V6T 1Z1, Canada}
\ead{djjones@physics.ubc.ca}

\begin{abstract}
We review the current state of tabletop extreme ultraviolet (XUV) sources based on high harmonic generation (HHG) in femtosecond enhancement cavities (fsEC).  Recent developments have enabled generation of high photon flux ($10^{14}$ photons/sec) in the XUV, at  high repetition rates ($>$50 MHz) and spanning the spectral region from 40 nm - 120 nm.  This level of performance has enabled precision spectroscopy with XUV frequency combs and promises further applications in XUV spectroscopic and photoemission studies.  We discuss the theory of operation and experimental details of the fsEC and XUV generation based on HHG, including current technical challenges to increasing the photon flux and maximum photon energy produced by this type of system. Current and future applications for these sources are also discussed.
\end{abstract}
\submitto{\JPB}

\section{Introduction}
\label{intro}
The use of amplified femtosecond (fs) optical pulse trains to drive high harmonic generation (HHG) for creation of temporally and spatially coherent vacuum ultraviolet, or VUV (120 - 200 nm), extreme ultraviolet, or XUV (10 - 120 nm), and soft x-ray radiation (0.1 - 10 nm), is now well into its third decade. Over the years, this approach has achieved several important technical milestones \cite{Brabec2000}.  More recently, careful quasi-phase matching of the harmonics with the optical driving field has enabled production of radiation down to 4.4 nm or 282 eV \cite{Gibson2003}. With the advent of carrier-envelope phase control of fs pulses \cite{DJJones2000}, studies of attosecond physics with HHG-generated XUV pulses are now possible \cite{Baltuska2003}. Beyond these technical demonstrations is an extensive list of experimental  accomplishments: lensless imaging \cite{Sandberg2007}, imaging of valence electron motion \cite{Goulielmakis2010} and electron tunneling \cite{Uiberacker2007}, x-ray Fourier transform holography \cite{Sandberg2009}, and probing multi-electron dynamics \cite{Shiner2011}.
 
The driving fields needed for HHG have peak intensities $>10^{13}$ W/cm$^2$ and are normally generated using chirped pulse amplifier (CPA) solid-state laser systems in which a pulse selector first decimates the pulse train from a mode-locked oscillator (originally at 50-100 MHz) to a much lower repetition rate (1 Hz to 1 MHz) before stretching, amplifying, and recompressing the resulting high energy pulses (up to 1 J at 1 Hz). With the relatively long time between successive amplified pulses, the typical HHG setup can be considered as a single pulse interacting with a gas target: a single pass of the optical driving pulse through a gas target (normally either a gas jet or pressurized hollow fiber) results in a set of odd harmonics of the frequency of the fundamental field. In the time domain this is generally expressed as discrete and finite bursts of attosecond XUV pulses, and these bursts are generated at the repetition rate of the driving field.

For a number of reasons there has been a long-standing motivation to increase the repetition rate of HHG sources to the tens or hundreds of MHz. With a given (adequate) photon flux, a low repetition rate means the XUV pulses can have a large enough peak power that nonlinear effects in the XUV can corrupt experimental results. For example, in angle-resolved photoemission spectroscopy (ARPES) \cite{Damascelli2003}, if too many photoelectrons are created at the same time (by the same burst of XUV pulses), the resulting space charge will distort the liberated electron trajectories and lead to erroneous results. Indeed, these space charge effects have limited the widespread and effective use of HHG sources for ARPES experiments although some successes have been reported \cite{Mathias2007,Berntsen2011}.  In addition, low yield photoelectron experiments such as ARPES or COLTRIMS \cite{Ullrich2003} mean that prohibitive long data acquisition times ($>$10 Hours) are required for statistically significant data. Finally, of more recent interest is the goal of generating an XUV femtosecond frequency comb (FFC) to extend the revolution in precision spectroscopy now enabled by FFCs \cite{YeCundiffBook_2005} in the visible and near IR into the XUV spectral regions.  

A small sampling of the opportunities for applications of VUV and XUV frequency comb spectroscopy includes precision spectroscopy of hydrogen, and hydrogen-like helium ions in the XUV as fundamental tests of quantum electrodynamics \cite{Eyler2007}, photoionization of molecular Rydberg states for photochemical studies of atmospheric processes, and new optical clocks in the VUV frequency range.  An exciting new candidate for an optical clock transition involves an M1 isomeric nuclear transition in $^{229}$Th \cite{Hehlen2011}.  This unusually low lying nuclear state transition  is the only known nuclear transition of this type accessible by optical spectroscopy, and has an expected wavelength between  145 nm - 175 nm.  It has a natural linewidth of 10 mHz, and is expected to broaden to only 10 kHz when doped in an appropriate crystal host.  This application and other direct frequency comb spectroscopy experiments require repetition rates $>$100 MHz and generation of phase coherent, low noise combs in the XUV with as much power per comb element as possible.
 
Most CPA systems used in HHG are centered near 800 nm and are based on Ti:Sapphire as the gain medium. While these systems can produce the shortest (few-cycle) pulses, fundamental limitations in pump laser power and thermal management issues have limited the average output power without too much room for improvement.  This, in turn, limits the repetition rate as the threshold intensity of $>$10$^{13}$ W/cm$^2$ must still be met for HHG. As an example, a 10-fs pulse train at 100 MHz focused to a spot size (diameter) of 20 $\mu$m requires an average power of 30 W to even reach this lower limit. Alternatively, the higher conversion efficiency in Yb-doped glass and solid state systems along with high power, multi-mode pump laser diodes offers another route toward significantly higher average power \cite{Russbueldt2010}. However, it is currently difficult to maintain pulse widths much below 150 fs in these high power systems. Despite these challenges, a recent demonstration of XUV generation at 20 MHz in a single-pass arrangement was reported  using a Yb-doped solid state oscillator (Yb:KGW) followed by a Yb:YAG amplifier \cite{Vernaleken2011}. In this work, 74 fs pulses were produced using external spectral broadening and subsequent pulse compression.
 
\begin{figure}[h!]
\centering
\resizebox{0.75\textwidth}{!}{%
  \includegraphics{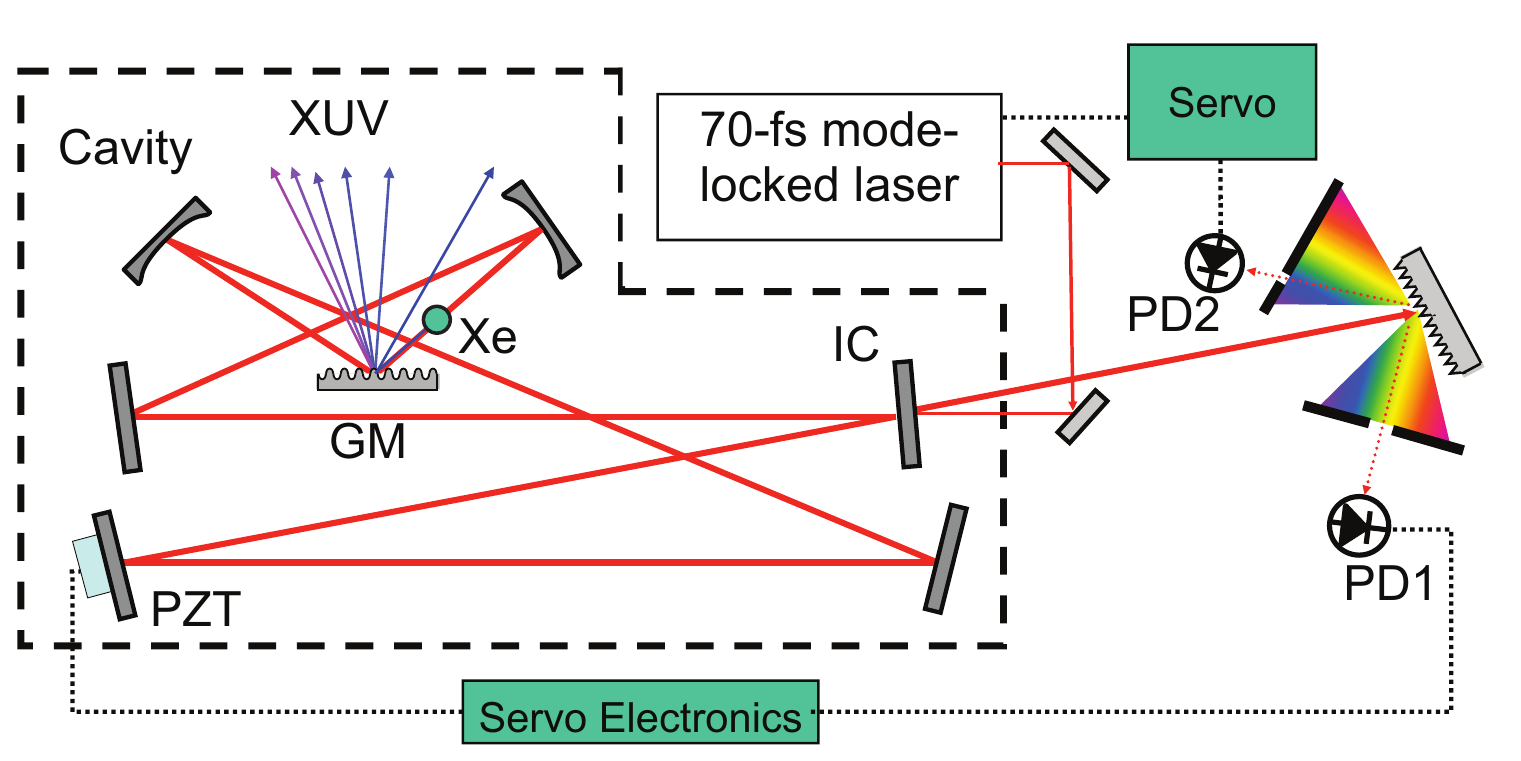}}
\caption{Schematic of a Ti:Sapphire fsEC-based XUV source using a diffraction grating mirror (GM) XUV output coupler and a xenon gas (Xe) target. Also labelled in the figure is the piezo transducer (PZT) used to control the cavity length, the fsEC input coupler (IC), and two photodetectors (PD1 and PD2) used in the fsEC stabilization.}
\label{fig:fsECSchematic}       
\end{figure}

An alternative method capable of achieving significantly higher repetition rates with suitable pulse energies uses a femtosecond enhancement cavity (fsEC) \cite{Jones2005,Gohle2005}, illustrated in \fref{fig:fsECSchematic}. With this technique, the output of a mode-locked oscillator is coupled to a high-finesse cavity.  Soon after full control of the carrier-envelope phase was realized in 2000, the feasibility of femtosecond pulse enhancement inside of optical resonators was considered \cite{Jones2002}.  By controlling the phase evolution of successive pulses incident on the cavity such that they phase-coherently add with the existing pulse stored in the cavity, significant enhancement of intra-cavity pulse energy can be realized at the full repetition rate of the mode-locked oscillator. A gas target is placed inside the cavity and these high energy, high repetition rate pulse trains drive the HHG process.

We begin this review with an overview of the operating principles behind a fsEC-based XUV source, including resonator fundamentals, the basics of HHG used to generate the XUV in the tight-focus regime of the fsEC, and the output couplers used to extract the harmonics from the cavity for use in experiments.  We then review experimental aspects of fsEC-based XUV generation, such as the fs oscillators used to seed the fsEC, the vacuum apparatus, active stabilization of the fsEC to the oscillator, delivery of the target gas to the intracavity focus, and finally detection of the harmonics coupled out of the cavity.  We proceed with a detailed overview of the approaches used to solve two of the greatest challenges that have been encountered in attempts to increase the XUV flux generated by these systems.  These two challenges are: 1) degradation of the very low loss cavity optics used to obtain large enhancement of the intracavity pulses, and 2) limitations imposed by the plasma generated at the intracavity focus.  We proceed with a description of four classifications of fsEC-based XUV sources developed to date, before moving on to discuss current and future applications of VUV and XUV frequency combs using the fsEC sources.  Central to the experiments conducted thus far has been establishing the coherence of the frequency comb and the limits on the frequency stability of its comb elements.  Finally, we conclude with comments on the outlook for further developments of the fsEC XUV sources.

\section{Theory of Operation}
\label{sec:fsEC_TechDetails}
In this section, we review the theoretical aspects surrounding fsEC-based XUV sources. We first review the fundamentals of an optical resonator with regard to a continuous wave (CW) single frequency optical input. We then extend this model to fsEC resonators operating at low intensity (where nonlinear effects do not occur). Within this linear regime the fsEC is simply an optical resonator whose (multiple) resonances are simultaneously excited by a comb of frequencies from a mode-locked laser oscillator. A simple and intuitive way to visualize the fsEC resonances is presented, which also illustrates the effect of  dispersion in the cavity.  We discuss the requirements for dispersion management in fsEC systems and several methods used to measuring it.  We proceed with an outline of the theory of HHG in the tight-focus regime, where the HHG occurs in a gas target placed near the tight focus of the fsEC.  We conclude this section with an overview of the methods currently used and recently proposed to efficiently couple the XUV out of the fsEC for XUV spectroscopic experiments.

\subsection{Frequency Comb Excitation of Optical Resonators}
\label{subsec:fsOptRes}
It is important to develop an intuition about how an optical resonator responds to pulsed excitation of a mode-locked laser, and when maximum intra-cavity buildup occurs.  In the time domain, one can imagine that the envelope of the pulse circulating in the fsEC must overlap with that of each subsequent pulse input from the oscillator.  A second requirement is that the optical carrier under the envelope must also constructively interfere. In other words, the round-trip carrier-envelope phase shift introduced by the cavity must match the pulse-to-pulse carrier-envelope phase shift from the oscillator.  In the frequency domain, these two requirements correspond to matching the pulse repetition rate $\frep$ of the mode-locked oscillator \cite{DJJones2000} to the fsEC and setting equal the relative carrier-envelope-offset frequencies of the fsEC and oscillator, that is $\Delta \fceo=0$.  Practically speaking, active stabilization of $\frep$ and possibly $\Delta \fceo$ is required, as is described in \sref{fsEC:Stabilization}.  The theory of optical resonators is well known, and what follows in this section is a brief summary of results that we find particularly relevant for fsEC work.  Similar material has also been presented in recent reviews of cavity-enhanced direct frequency comb spectroscopy \cite{Thorpe2008,Adler2010}.

\begin{figure}[h!]
\centering
\resizebox{0.75\textwidth}{!}{%
  \includegraphics{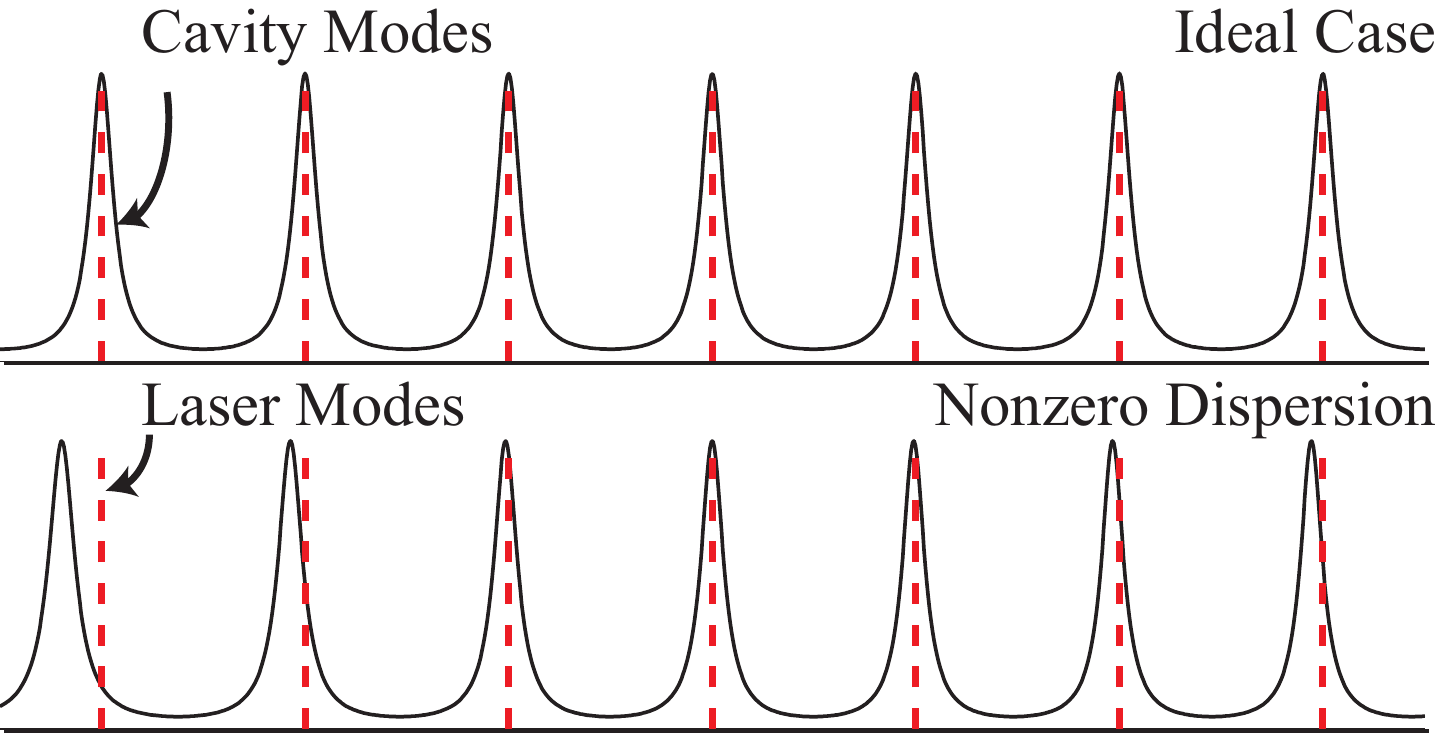}}
\caption{Illustration of the effect of dispersion on the alignment of cavity modes and mode-locked laser modes for the ideal, dispersion free case (top), and a non-zero net cavity dispersion (bottom), in which cavity modes have an unequal spacing across the optical frequency spectrum of the mode-locked laser. The amount of dispersion is greatly exaggerated for illustration purposes.}
\label{fig:DispersionCartoon}       
\end{figure}

Physical insight can be gained by examining the resonance condition of an fsEC excited by the mode-locked laser optical spectrum, or frequency comb.  To approach this problem we begin with the general resonance condition of an optical resonator,
\begin{equation}
\label{eq:ResCondition}
2\pi m = \frac{\omega}{c}L+\Phi(\omega),
\end{equation}
which requires that the phase accumulated by an optical wave of frequency $\omega$ in one round trip of a resonator of length $L$ is equal to $2\pi m$, where $m$ is an integer, and c is the speed of light in vacuum.  In addition to the phase shift due to propagation in free space, an additional frequency-dependent phase $\Phi(\omega)$, or dispersion, is added by elements within the cavity.  Dispersion leads to a nonuniform spacing (as a function of frequency) of the resonances, which cannot be exactly aligned with the equal spacing of the modes produced by the mode-locked oscillator, as depicted in \fref{fig:DispersionCartoon}.  Excessive dispersion leads to a reduction in bandwidth and longer pulses, and the amount of filtering depends on the finesse of the cavity and the spectral bandwidth of the mode-locked oscillator. In a bare fsEC in vacuum, $\Phi(\omega)$ is solely due to the dielectric mirrors that make up the resonator, but when a gas target is present, additional terms arise to account for the dispersion of the gas and the plasma generated.

Following the standard analysis of a Fabry-Perot resonator, the steady state intensity circulating in the cavity is given by,

\begin{equation}
\label{Eq:CavityIntensity}
I_c(\omega,d)=\frac{t^2_iI_0(\omega)}{(1-r_ir_{cav})^2+4r_ir_{cav}\sin^2(\phi/2)},
\end{equation}
where $I_0(\omega)$ is the incident intensity, $t_i$ and $r_i$ are the magnitude of the field transmission and reflection coefficients of the input coupler, respectively, and $r_{cav}$ is the magnitude of the field reflection coefficient for the remaining cavity mirrors, combined. The field and intensity reflection coefficients are related by, $R_i=r_i^2$, and likewise, $T_i=t_i^2$. Generally, very low loss mirrors are available such that  $R+T\cong1$, for visible and near-infrared optical frequencies.  The resonance condition information is included in the phase, $\phi=\frac{\omega}{c}d+\Phi(\omega)$. Here we have written only the incident electric field and the cavity phase explicitly as functions of frequency, but strictly speaking the mirror reflectivity and transmission are also functions of optical frequency.

\Eref{Eq:CavityIntensity} can be used to derive an expression for the cavity finesse, $\mathcal{F}$, which is defined as the ratio of the cavity free spectral range (FSR) to the cavity full width at half maximum (FWHM).  For sufficiently large values ($\mathcal{F}\gg1$), the cavity finesse is given by,

\begin{equation}
\mathcal{F}=\frac{\pi\sqrt{r_ir_{cav}}}{1-r_ir_{cav}}.
\end{equation}

The cavity enhancement $B(\omega,d)={I_{c}(\omega,d)}/{I_{0}(\omega)}$ of the intensity can also be expressed in terms of the finesse and the resonance condition phase,

\begin{equation}
\label{Eq:Buildup}
B(\omega,d) = \left(\frac{t_{i}}{1-r_{i}r_{cav}}\right)^{2}\frac{1}{1+\left(\frac{2\mathcal{F}}{\pi}\right)^2\sin^2(\phi/2)}.
\end{equation}

The maximum enhancement is achieved when the input coupler intensity transmission is equal to all other losses in the cavity, and in this case the resonator is said to be impedance matched.  In this limit, the enhancement factor is approximately $\mathcal{F}/\pi$. When there is a lossy element in the cavity, such as in intracavity second harmonic generation, impedance matching can be particularly important. From \eref{Eq:Buildup}, one can see that fluctuations in the intracavity phase, $\phi$, are magnified by a factor of $\mathcal{F}^2$.  As a result, for intracavity HHG in the fsEC, it is often desirable to limit or spoil the finesse by increasing the input coupler transmission.  When the finesse is reduced in this way, and the net cavity loss is dominated by the input coupler, the cavity enhancement is $B\approx2\mathcal{F}/\pi$.      In summary, in these two cases the cavity enhancement is,
\numparts
\begin{eqnarray}
B_{max}(\omega,d) &\approx& \frac{\mathcal{F}}{\pi} \hspace{0.5cm} \mathrm{(Impedance~matched)} \\
 &\approx& \frac{2\mathcal{F}}{\pi} \hspace{0.35cm} \mathrm{(Input~coupler~limited)}.
\end{eqnarray}
\endnumparts

The expressions presented in this section are derived for one single-frequency mode resonating in the cavity and typically the frequency dependence is not explicitly written.  As long as nonlinearities do not arise anywhere in the cavity we can simply apply this theory to the case of excitation with the frequency comb generated by a femtosecond mode-locked oscillator by keeping track of the resonance condition and enhancement for each comb element.  We expand this notion in the next section, and we discuss the nonlinear effects that arise when a plasma is generated in the cavity in \sref{sec:PlasmaEffects}.

\subsection{The fsEC Resonance Map}
We have found it useful to visualize the behavior of the fsEC using a \emph{resonance~map}, which is a plot of the cavity enhancement $B(\omega,d)$.  For a CW laser source, a single row (column) of this map is the well-known Fabry-Perot resonance structure that is observed when the cavity length (laser frequency) is scanned.  The situation is slightly more complicated when the cavity is simultaneously excited with $\sim10^5$ modes typical in a mode-locked oscillator; in this case it is easiest to compute the cavity enhancement numerically and plot the enhancement on a heat map type plot.

\begin{figure}[h!]
\centering
\resizebox{0.75\textwidth}{!}{%
  \includegraphics{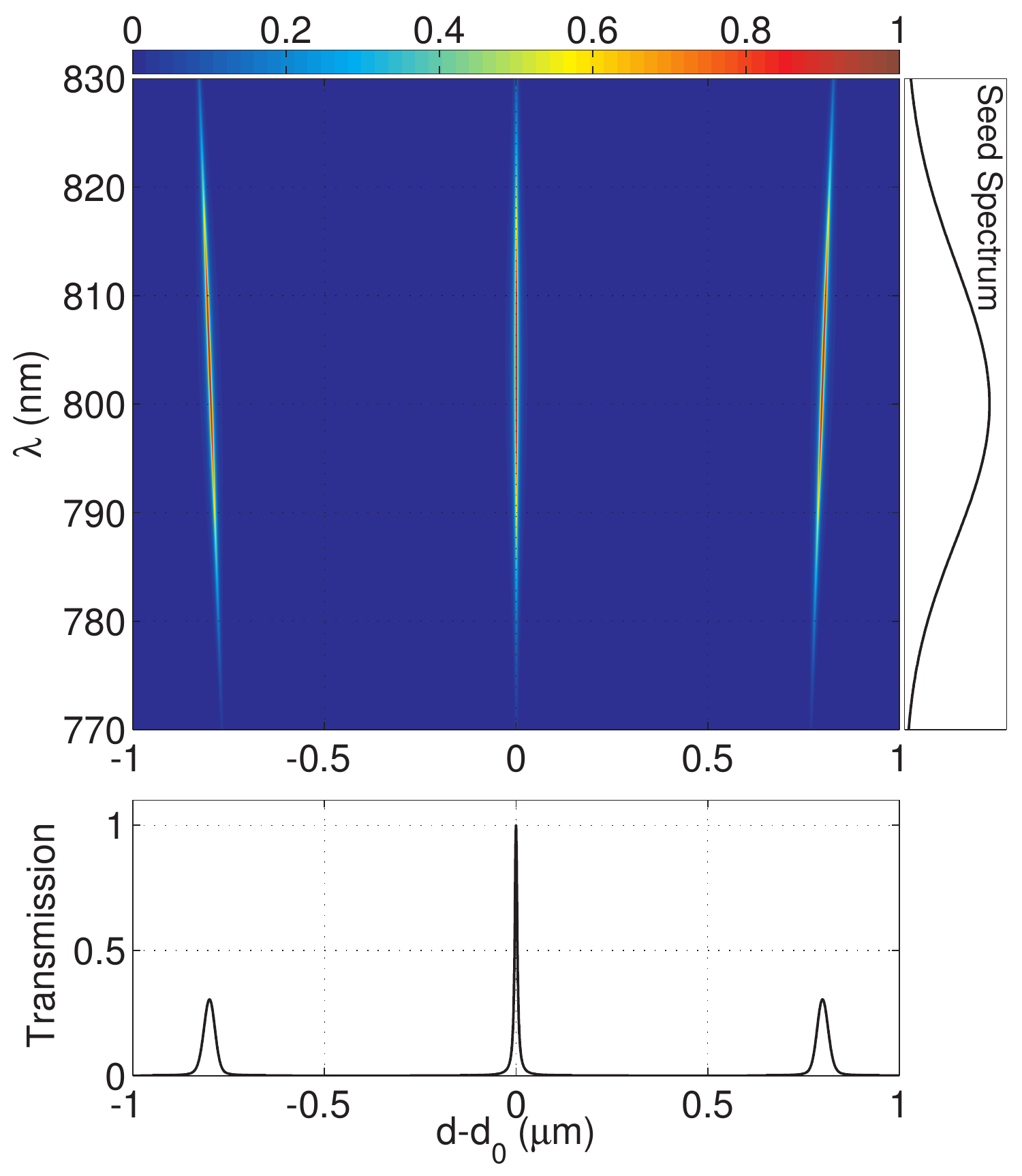}}
\caption{Resonance map (top) for a fsEC with $\mathcal{F}=50$, $\fceo=0$, and $\Phi=0$, excited with an incident pulse centered at 800 nm with a 30 nm spectral bandwidth (right), and the resulting transmission (bottom) when the cavity length is scanned near the \emph{length-matched} central fringe, $d=d_0$. }
\label{fig:ResonanceMap}       
\end{figure}

The seed pulse train is taken to be a frequency comb with modes separated by the repetition rate of the oscillator, with a spectral envelope function to simulate the bandwidth of the oscillator.  We then compute the cavity intensity on a 2D grid representing the resonance map.  We show an example resonance map in the top panel of \fref{fig:ResonanceMap} for the case of zero cavity dispersion, $\Phi(\omega)=0$, the relative carrier-envelope frequency difference between the oscillator and cavity is zero, a spectral bandwidth of 30 nm centered at 800 nm, and relatively low finesse, $\mathcal{F}=50$, to introduce its utility. Resolving the individual comb elements, which are separated by several tens to hundreds of MHz, is not necessary for the resonance map.  On the other hand, we do require high resolution for the cavity length axis.  We plot three cavity resonances in \fref{fig:ResonanceMap}, spanning a cavity length range of about 2 $\mu$m.  We see that there is only one cavity length, $d=d_0$, where all modes of the input spectrum are resonant simultaneously in the cavity.   This \emph{central fringe} has the largest enhancement, $B$, and occurs at the cavity length where the round-trip time equals the oscillator pulse period.   The adjacent fringes ($d-d_0=\pm\lambda$) are \emph{tilted} because the resonance condition is not satisfied for all frequency comb modes at the same cavity length simultaneously.
The bottom panel of \fref{fig:ResonanceMap} represents the cavity transmission as a function of the cavity length, where the transmitted intensity, at a given cavity length, is proportional to the sum of the intensity of all frequency modes stored in the fsEC. The amplitude of the modes adjacent to the central fringe are considerably reduced due to the tilt of the corresponding fringes in the resonance map.

\begin{figure}[h!]
\centering
\resizebox{0.75\textwidth}{!}{%
  \includegraphics{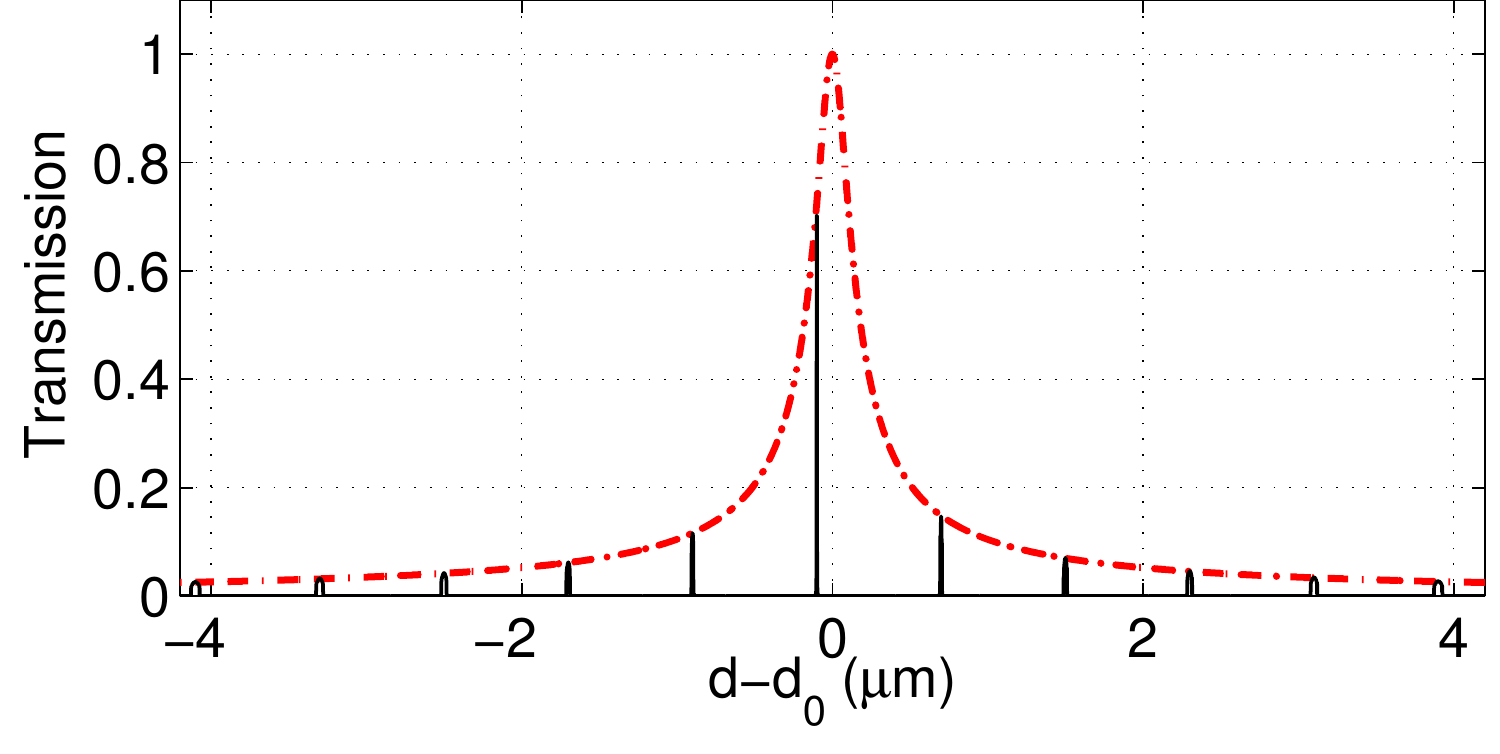}}
\caption{Cavity transmission for $\mathcal{F}=500$, $\fceo=\frep/8$, and $\Phi=0$, when the cavity length is scanned near the \emph{length-matched} central fringe ($d=d_0$).  The maximum amplitude of transmission fringes is indicated by the envelope (dotted line).}
\label{fig:CavityTransmission2}       
\end{figure}

\Fref{fig:CavityTransmission2} illustrates how the cavity transmission is modified when the offset frequency, $\fceo$, is varied.  In general the cavity contributes its own carrier-envelope phase and the relevant parameter for the fsEC is the relative difference, $\Delta\fceo$, between the oscillator $\fceo$ and the carrier-envelope offset of the cavity. \Fref{fig:CavityTransmission2} shows several cavity transmission fringes for a non-zero relative offset frequency, $\Delta \fceo$, of the incident field for a cavity finesse, $\mathcal{F}=500$, and $\Delta \fceo=\frep/8$, and still no cavity dispersion, $\Phi(\omega)=0$.  We see the cavity resonances are shifted relative to the resonances for $\Delta\fceo=0$, and the peak of the central fringe does not occur at $d=d_0$.  In fact, as $\Delta \fceo$ is varied from zero, the peak of the fringes map out an envelope function, shown as a dotted red line that reaches a maximum at $d=d_0$.    The peak of the envelope function corresponds to $\Delta\fceo=0$. This envelope function has been described previously \cite{Jones2000,Arissian2009}.  The narrow width of the envelope (dotted line in \fref{fig:CavityTransmission2}) illustrates the limited range over which the cavity length and offset frequency can vary while maintaining maximum transmission (and therefore maximum cavity enhancement, $B$). Increasing the optical bandwidth or the cavity finesse further narrows this range.  

\begin{figure}[h!]
\centering
\resizebox{0.75\textwidth}{!}{%
  \includegraphics{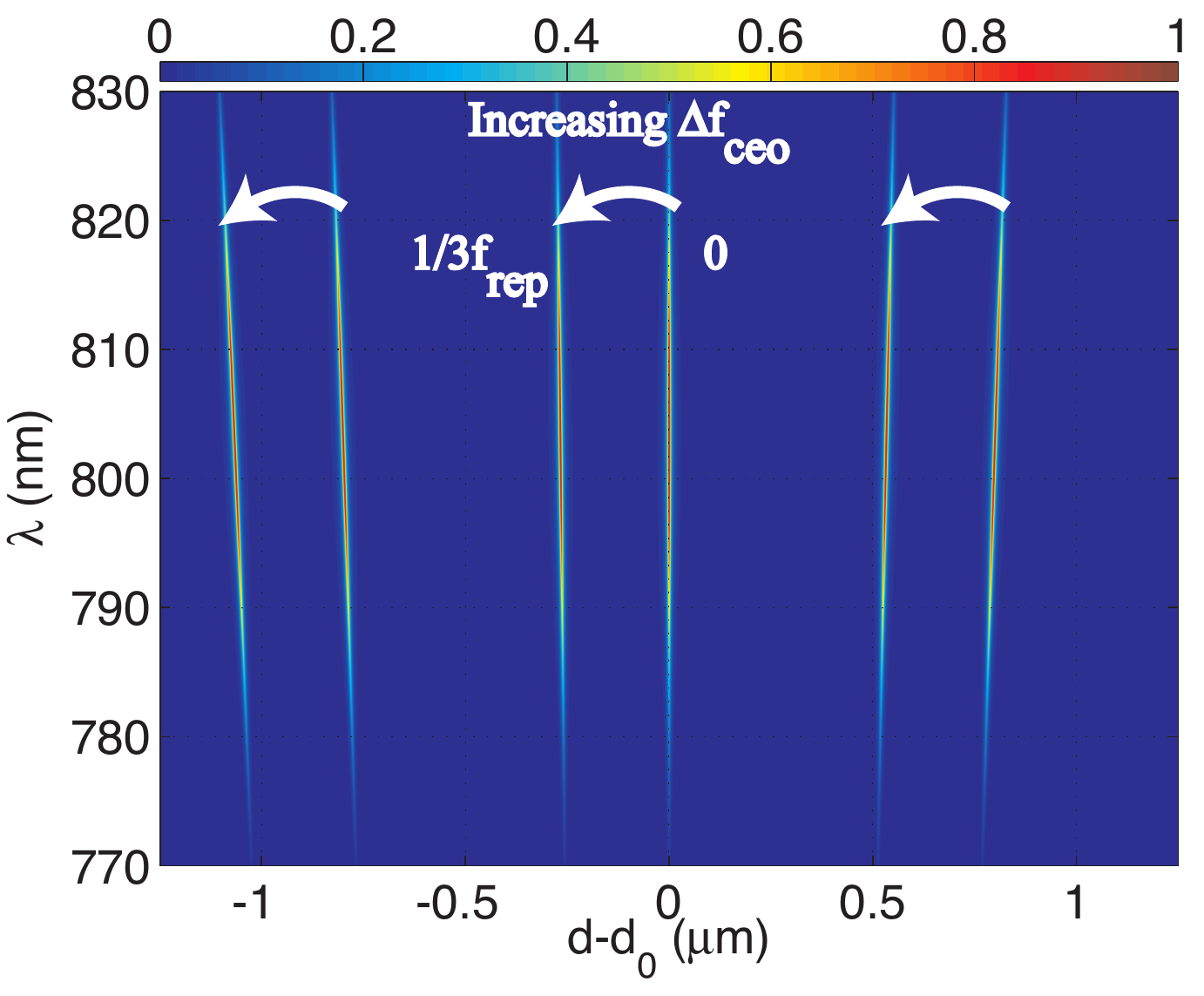}}
\caption{Resonance map for a fsEC with $\mathcal{F}=50$ illustrating the effect of varying the offset frequency difference, $\Delta\fceo$, between the fsEC and the oscillator.  As $\Delta\fceo$ is increased the resonance fringes move to the left as indicated.}
\label{fig:OffsetCartoon}       
\end{figure}

As the offset frequency is changed, the resonances as seen in the resonance map undergo a continuous evolution as well.  For example, as $\Delta \fceo$ is increased from its value of zero in \fref{fig:ResonanceMap}, the fringes shift from right to left and also change their tilt angle.  This is illustrated in a resonance map in \fref{fig:OffsetCartoon}, for two sets of fringes: 1) for $\Delta\fceo=0$, which has a central fringe centered at $d-d_0=0$ and 2)  for $\Delta\fceo=\frep/3$.  When $\Delta \fceo$ is further increased, the tilted mode that started near $d-d_0=800$ nm ends at $d=d_0$ perfectly vertically aligned when $\Delta\fceo=\frep$.

There is another feature of the resonance map that is interesting to note.  If one examines the behavior of cavity fringes much farther away from the central fringe ($|d-d_0|\gg\lambda$), it is seen that the tilt of the fringes becomes greater as $|d-d_0|$ increases.  Eventually, multiple fringes will have some frequency range resonant in the fsEC for the same cavity length.  For large $|d-d_0|$, ranging from one hundred $\mu$m to a few mm, the cavity transmission consists of a series of distinct transmission peaks, each corresponding to a cavity fringe coming into resonance in a different portion of the optical spectrum.  These fringes become increasingly more dense and lower in intensity as the cavity length moves farther from the location of the central fringe.  This is a very useful feature when performing the rough length-matching of the fsEC during the initial setup.  Of course, this behavior has been observed previously in the context of cavity-enhanced absorption spectroscopy \cite{Gherman2002}, but we point it out here to illustrate the power of the fsEC resonance map as a means of unifying the variety of behavior observed in the fsEC.

\subsection{Dispersion in the Femtosecond Enhancement Cavity}
\label{sec:Dispersion}
The resonance map is also a powerful tool to visualize the effect of dispersion, which is simply a variation in the resonance condition as a function of frequency, represented as $\Phi(\omega)$ in \eref{eq:ResCondition}.

\begin{figure}[h!]
\centering
\resizebox{0.75\textwidth}{!}{%
\includegraphics{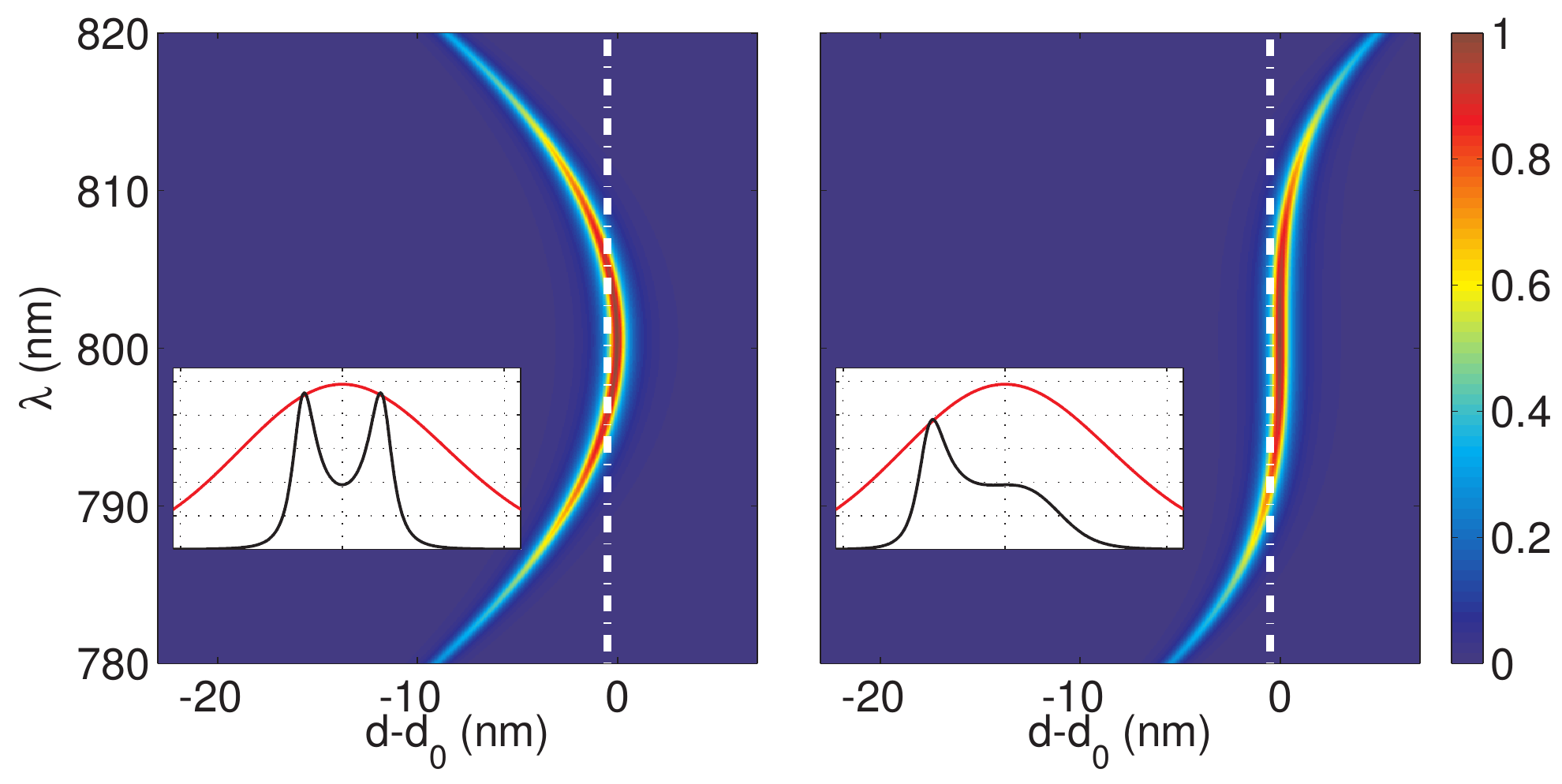}}
\caption{Resonance map of the central fringe (defined in text) for a fsEC with $\mathcal{F}=500$, $\fceo=0$, and an input spectral bandwidth of 30 nm, when 20 fs$^2$ of second order (left) and 200 fs$^3$ of third order (right) dispersion are added to the fsEC. The inset in each represents the spectrum that could be stored if the cavity were stabilized at the cavity length indicated by the dotted white line (\emph{i.e.} d-d$_0=-0.5$~nm), plotted with the incident spectrum of the seed.}
\label{fig:ResonanceMapDispersion}       
\end{figure}

For high finesse cavities ($\mathcal{F}>1000$) dispersion of the cavity mirrors will limit the enhancement if the residual group delay dispersion (GDD) is more than a few fs$^2$.  Such high finesse cavities prove to be difficult to work with for a number of reasons that will be discussed further in \sref{sec:fsEC_ Challenges}. The dispersion requirements for cavities with a finesse of a few hundred are significantly relaxed, but even modest second and third order dispersion can significantly restrict the spectrum supported by the fsEC.  In \fref{fig:ResonanceMapDispersion}, we plot the resonance map for two functional forms of the cavity phase: 1) $\Phi(\omega)$ is a quadratic spectral phase with a magnitude of 20 fs$^2$ (left), and 2) $\Phi(\omega)$ is a cubic spectral phase with a magnitude of 200 fs$^3$.  The inset on the left panel shows the spectrum that is stored in the cavity when the cavity length is stabilized to $d-d_0=-0.5$~nm.  This spectrum is a vertical slice of the resonance map, indicated with the dotted white line.

\begin{figure}[h!]
\centering
\resizebox{0.75\textwidth}{!}{%
  \includegraphics{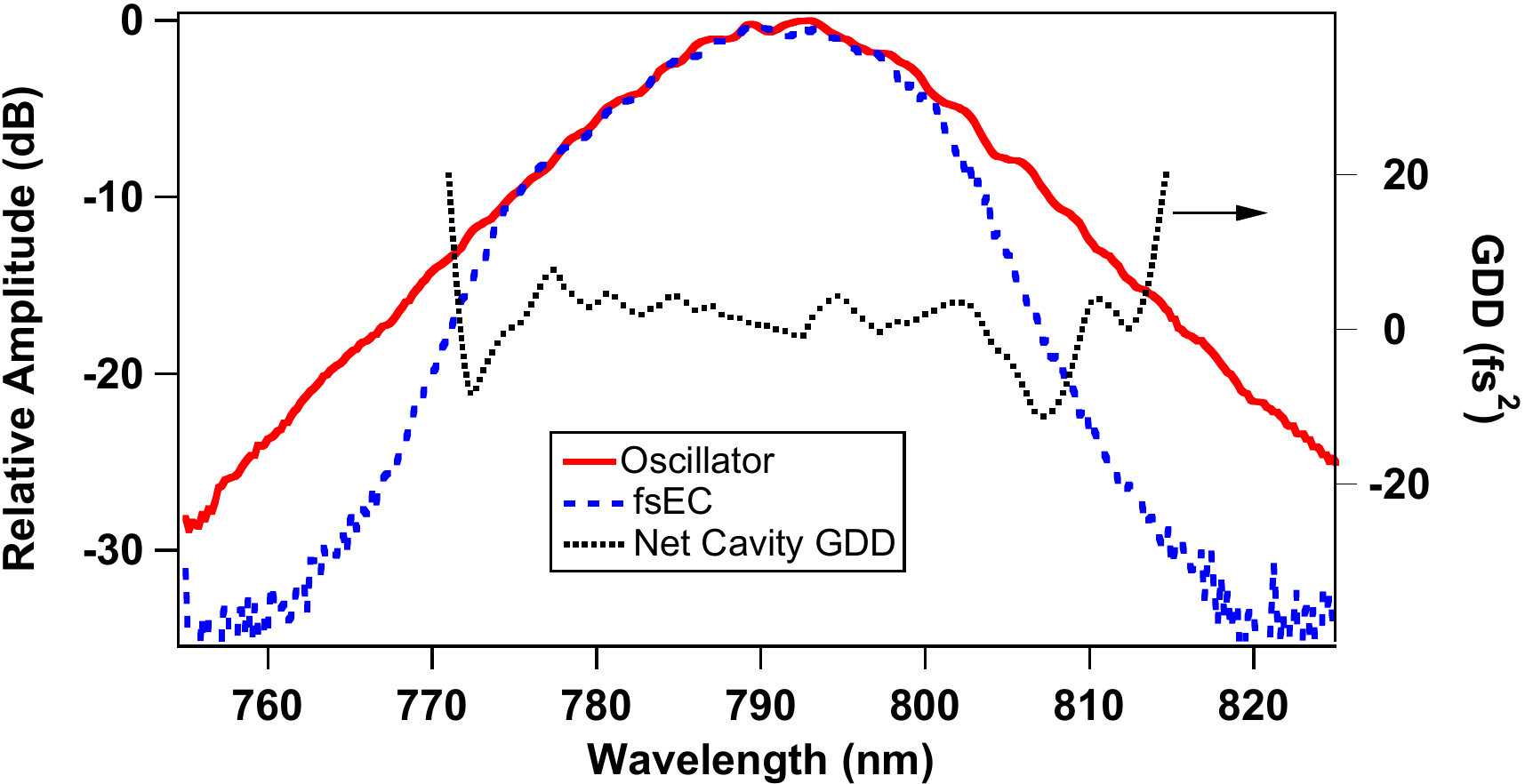}}
\caption{Oscillator and stored fsEC spectra for a Ti:Sapphire fsEC system with a high finesse ($\mathcal{F}>2000$) and low GDD cavity mirrors and a few fs$^2$ residual GDD. The GDD was measured using the technique of \cite{Hammond2009}.}
\label{fig:StoredSpectrum}       
\end{figure}

We show a set of experimental spectra for a pulse stored in the fsEC and the Ti:Sapphire seed oscillator in \fref{fig:StoredSpectrum}.  This figure shows that the majority of the optical spectrum can be enhanced in a high finesse cavity ($\mathcal{F}>2000$) with commercially available low GDD mirrors. Higher power seed sources enable lower-finesse cavities to be used, and therefore a further increase in the spectral bandwidth stored in the fsEC is possible.

Several similar methods of measuring the dispersion of the fsEC have been developed, each of which can be useful depending on technical details of the individual frequency comb source and fsEC combination.  Each of the schemes involves interrogating the spectral dependence of the resonance condition of \eref{eq:ResCondition} in one way or another.  The least technically demanding, if somewhat time consuming, method \cite{Hammond2009} involves scanning across the cavity resonances and measuring the differential change in the resonance cavity length, $\Delta d$, as a function of optical frequency relative to a reference optical frequency.  The repeatability and accuracy of this measurement depends on the stability of the oscillator's $\fceo$, which may need to be stabilized for the measurement of small cavity dispersion and high-finesse cavities.  Systems with a reasonably stable offset frequency (typically fiber lasers) do not need to be stabilized, but in either situation the numerical value of $\fceo$ is not needed for the dispersion measurement.

In another cavity dispersion measurement technique \cite{Schliesser2006}, the enhancement cavity is locked to the oscillator at a fixed, narrow band of optical frequencies and then $\fceo$ of the oscillator is varied.  For each value of $\fceo$, the optical spectrum stored in the enhancement cavity, which is a sensitive function of the cavity phase, $\Phi(\omega)$, is recorded.  This technique requires that the step size of the $\fceo$ scan be known, but its absolute value is not necessary. This method is very convenient when an acousto-optic modulator already exists in the system for other purposes, such as carrier-envelope offset frequency stabilization described in \sref{fsEC:Stabilization}.

Finally, in a third technique \cite{Thorpe2005} the average frequency of a narrow band of optical frequencies of the laser is locked to the cavity and then the repetition rate of the laser is scanned.  For each value of $\frep$ the optical spectrum stored in the cavity is recorded, and as in the previously described technique, the resulting 2D map depends on the cavity phase, $\Phi(\omega)$.

Each of the previous three techniques were developed to measure the dispersion characteristics of a bare cavity, and therefore requires linear behavior of the cavity to obtain this information.  It may also be desirable to study the dispersion properties of the cavity at higher intensities where nonlinearities are important, or for a nonlinear medium present in the cavity.  For this purpose spatial spectral interferometry (SSI) has been applied to fsECs \cite{Pupeza2010}.  This technique utilizes the interference between a reference beam and a second beam passed through a dispersive `device under test'  (\emph{i.e.} the fsEC) to determine the frequency dependent phase introduced by the fsEC.  A small angle between the two beams is introduced in one dimension (\emph{e.g.} vertical), while a grating is used to disperse the beams spectrally in the other direction (horizontal). The fringes that appear contain the spectral phase information desired, for example, parabolic shaped fringes appear for a purely parabolic spectral phase (GDD) added by the device under test. This measurement can be completed while the cavity is loaded, meaning that it is locked to the seed oscillator with high intracavity intensity, and is thus capable of measuring the dispersion resulting from nonlinear phenomena in the fsEC.

In summary, dispersion in the fsEC is a factor that must be considered carefully in optimizing a fsEC-based XUV source, and one must select low dispersion mirrors for the cavity.  Beyond the requirement of low GDD mirrors, the use of relatively low finesse cavities ($\mathcal{F}<500$) significantly reduces the necessity to manage dispersion.  More care must be taken to control the dispersion in the cavity if one wishes to increase the bandwidth supported by the cavity, use a higher finesse, or use a dispersive intracavity element (such as a Brewster plate for an XUV output coupler) 

\subsection{HHG in the Tight Focus Regime}
\label{Theory:PhaseMatching}
High harmonic generation is a well known nonlinear process that has experienced rapid, sustained development since the earliest experiments over two decades ago \cite{Brabec2000,Corkum1993}.  HHG is usually explained classically in terms of the three step model in which the Coulomb potential that binds the electron to the atomic nucleus is perturbed by an intense laser field.  If the external field is great enough, the electron can tunnel through the potential barrier whereupon it is accelerated away from the nucleus by the external laser field.  After a short time, the oscillating electric field reverses direction and accelerates the electron toward the nucleus.  If the nucleus and the electron recombine, a high energy photon is released with an energy equal to the sum of the ionization potential of the atom and the ponderomotive energy the electron acquires due to accelerating in the laser field.  In general, a series of odd harmonics of the fundamental laser frequency is produced.

Theoretical classical models and quantum mechanical calculations of the HHG process have been developed that very accurately describe the effects seen in experiments.  These calculations determine the intensity spectrum of the generated harmonics by single atoms and the phase of the radiators, which must be incorporated to predict the harmonic emission from a macroscopic ensemble of atoms. An integral part of the picture is the continuum dynamics of the electron after it is released from the atom.  Classical models reveal that above a certain threshold intensity two pathways are possible, a long and a short trajectory, corresponding to the extent to which the electron travels in the continuum.  Quantum mechanically, an electron can take many pathways through the continuum.  For high intensities such that the Coulomb potential is negligible compared to the electron energy gained by the acceleration of the laser field (the ponderomotive energy), the quantum pathways follow closely the classical trajectories in generating the high harmonics.  Interestingly, harmonics with photon energy near the ionization potential also show multiple pathways when the Coulomb potential is included in the model \cite{Hostetter2010}.   The case of HHG in a fsEC currently falls into this latter category, and optimizing harmonic generation in a fsEC inevitably relies on developing an understanding of the HHG process for the conditions present near the tight focus of the fsEC.
 
In nearly all optical nonlinear processes, including HHG, the phase evolution between the fundamental and generated field is a key parameter. Addressing the phase mismatch between the fundamental and HHG fields is necessary to optimize the XUV flux. Shortly after the discovery of HHG, it was realized that the focus geometry played a large role in the efficiency of the harmonic conversion process \cite{Li1989,LHuillier1991}. In the tight-focus regime, it was found that the harmonic flux scales as $b^3$ for a fixed intensity \cite{Lompre1990}, where $b$ is the confocal parameter of the focused Gaussian beam. Improved technology enabled further investigation into the loose focus regime and HHG within gas-filled hollow fiber waveguides \cite{Rundquist1998,Constant1999,Durfee1999}. Through better understanding of the waveguide and plasma dispersion it became possible to incorporate these parameters into a phase matching procedure, and thereby improving the yield and beam shape of certain harmonics. The use of waveguides was further revolutionized through quasi-phase matching by modulating the diameter of the waveguide \cite{Paul2003}. This improvement enables phase matching up to several hundred eV \cite{Gibson2003}. Furthermore, counterpropagating pulses have been used in the loose focus geometry \cite{Peatross1997,Voronov2001} to turn off the generation of harmonics during phase mismatched events.  This idea was expanded to gas-filled waveguides \cite{Cohen2007,Zhang2007}, where the effect of counterpropagating pulse trains were used to further increase the XUV flux through improved phase matching.

There is likewise a push to maximize the harmonic flux generated in fsECs. In this section we present an overview and optimization of the HHG process via a numerical model, while keeping in mind the constraints imposed by the fsEC. In order to reach the necessary peak intensity, fsEC HHG systems are required to have a tight focus.  In this regime the Gouy phase experienced by the driving fundamental field severely limits phase matching. Other factors that are requisite to fsECs include the fact that the fundamental field is very sensitive to strong nonlinear phase shifts (such as due to plasma creation) and loss (as in the case of hollow-fiber coupling). These factors highlight the need to optimize the gas nozzle geometry in fsEC systems, so as to maximize the harmonic generation from the minimum required amount of gas/plasma.

In fsECs the optimal gas nozzle geometry concentrates the gas target to a much smaller interaction region, and at much higher pressures than used in some previous studies. Nozzles used in these systems produced relatively long interaction lengths and to keep plasma phase shifts small, low gas pressure was used.  We address effects of different gas nozzle designs in \sref{sec:GasTarget}. In this low-pressure, long-interaction-length regime the phase matching can vary significantly for different harmonic orders, particularly for near-threshold harmonics that are generated with fsECs. On the other hand the optimal harmonic flux requires high pressure and tight confinement of the gas to a region very close to the peak of the intracavity intensity. To study both of these regimes, we applied previous theoretical work to develop an on-axis model of the harmonic generation process for the harmonic orders produced in fsEC XUV sources \cite{Hammond2011}.  In what follows we briefly review some of the key points of the theory here, but full details of the numerical model are provided elsewhere \cite{Hammond2011,Hammond2011a}.

To develop a numerical model we first consider the amplitude of the harmonics generated on-axis.  The field amplitude of the $q^{th}$ harmonic $E_q$ can be found by solving,
\begin{eqnarray}
\frac{\partial}{\partial z} E_q(z) =-\frac{\rho(z) \sigma(\omega_{q})}{2} E_q(z) + \frac{i \mu_0 q^2 \omega_1^2}{2 k_q} \rho(z) |\tilde{d}(\omega_q,z)| e^{i \Delta\phi(z)},
\end{eqnarray}
where the gas density is $\rho(z)$ and $\sigma(\omega_{q})$ is the absorption cross-section of the $q^{th}$ harmonic. The dipole response, $\tilde{d}(\omega_q,z)$, is calculated by solving the time dependent Schr\"{o}dinger equation.  In \cite{Hammond2011} the dipole response is not deconstructed into magnitude and phase contributions from separate quantum trajectories directly from the calculated dipole moment, as discussed elsewhere \cite{Yost2009,Gaarde2008}.  Instead, long and short trajectory phases are added phenomenologically as described below. This approach has its limitations; for example it would not be able to compute the relative contributions to the harmonic intensity from long and short trajectories, nor would it predict the interference that arises between them.  However, the model does capture much of the physics we are interested in, primarily the variation of the generated harmonics with the geometry and density of the target gas for a given intensity within the fsEC.

The phase mismatch is $\Delta \phi(z)=q \phi_{1}-\phi_{q}$ with,
\numparts
\begin{eqnarray}
q \phi_1(z) &=& q k_1 n(z,\omega_1) z-q\tan^{-1}\left(\frac{z}{n(z,\omega_1)z_R}\right)+ \Phi(z)  \\
\phi_q(z) &=& k_q n(z,\omega_q) z - \tan^{-1}\left(\frac{z}{n(z,\omega_q) z_R} \right),
\end{eqnarray}
\endnumparts
where $z_{R}=b/2$ is the Raleigh range of the fundamental beam ($z_{R}\sim 0.5$~mm in tight-focus fsEC systems). The phase of the atomic response is given by $\Phi(z)=-\alpha_iI(z)$, where $\alpha_{i}$ parameterizes the different trajectories.   In \cite{Hammond2011} the short and long trajectory phase coefficients values of $\alpha_i = 1\times10^{-14}$ and $25\times10^{-14}$ cm$^2$/W were used, respectively, consistent with the classical phases of the electron for 800 nm excitation.  The model allows us to calculate the effect of different gas jet geometries on the harmonic generation process through the refractive index $n(z,\omega_q)$, which is related to the gas density and plasma by,
\begin{eqnarray}
n(z,\omega_{q}) &=& 1 + \frac{\rho(z)}{\rho_0} \left[\left(1-\eta \right)\delta_{q} - \frac{\omega_p^2}{2 \omega_{q}^2} \eta \right],
\label{eqn:index}
\end{eqnarray}
where $\rho_0$ is the gas density at 1 ATM, $\eta$ is the ionization fraction, $\delta_{q}$ is the xenon index correction from vacuum, and $\omega_p$ is the plasma frequency.  The ionization fraction includes a dynamic ionization component that changes during each pulse and a steady state component that arises due to the accumulation of plasma resulting from multiple pulses.  The ionization fraction  is estimated by combining calculations for the ionization rate \cite{Yudin2001} with estimates of the transit time of atoms through the interaction region based on gas velocity and plasma expansion dynamics. Different gas jets will lead to very different ionization fraction, which affects the harmonic yield and beam shape of the generated harmonics.  We discuss these effects in \sref{sec:GasTarget}.

The results of these simulations are consistent with previous work on above-threshold harmonics in the tight-focus regime with respect to fundamental intensity and gas density \cite{LHuillier1992PRA,Salieres1995,Altucci1996}. A notable difference exists in the higher intensity regime used in these studies, in that the high pulse energy leads to a significantly higher ionization fraction at the focus. This manifests itself in a quantitative difference in the harmonic yield for different gas nozzle positions relative to the focus. Thus, in the tight-focus regime within a fsEC the optimal position for the HHG process occurs much nearer to the focus with a very high gas density. Interestingly, the tight-focus regime is also relevant in recent development of high repetition rate single-pass HHG systems operating near 100 kHz with several $\mu$J pulses.  A very recent work \cite{Heyl2012} investigates the role of phase matching in this regime, and considers how the phase matching is modified when the focal length of a lens used to achieve the tight focus is used as a scaling parameter.

\begin{figure}[h!]
\centering
\resizebox{0.75\textwidth}{!}{
 \includegraphics{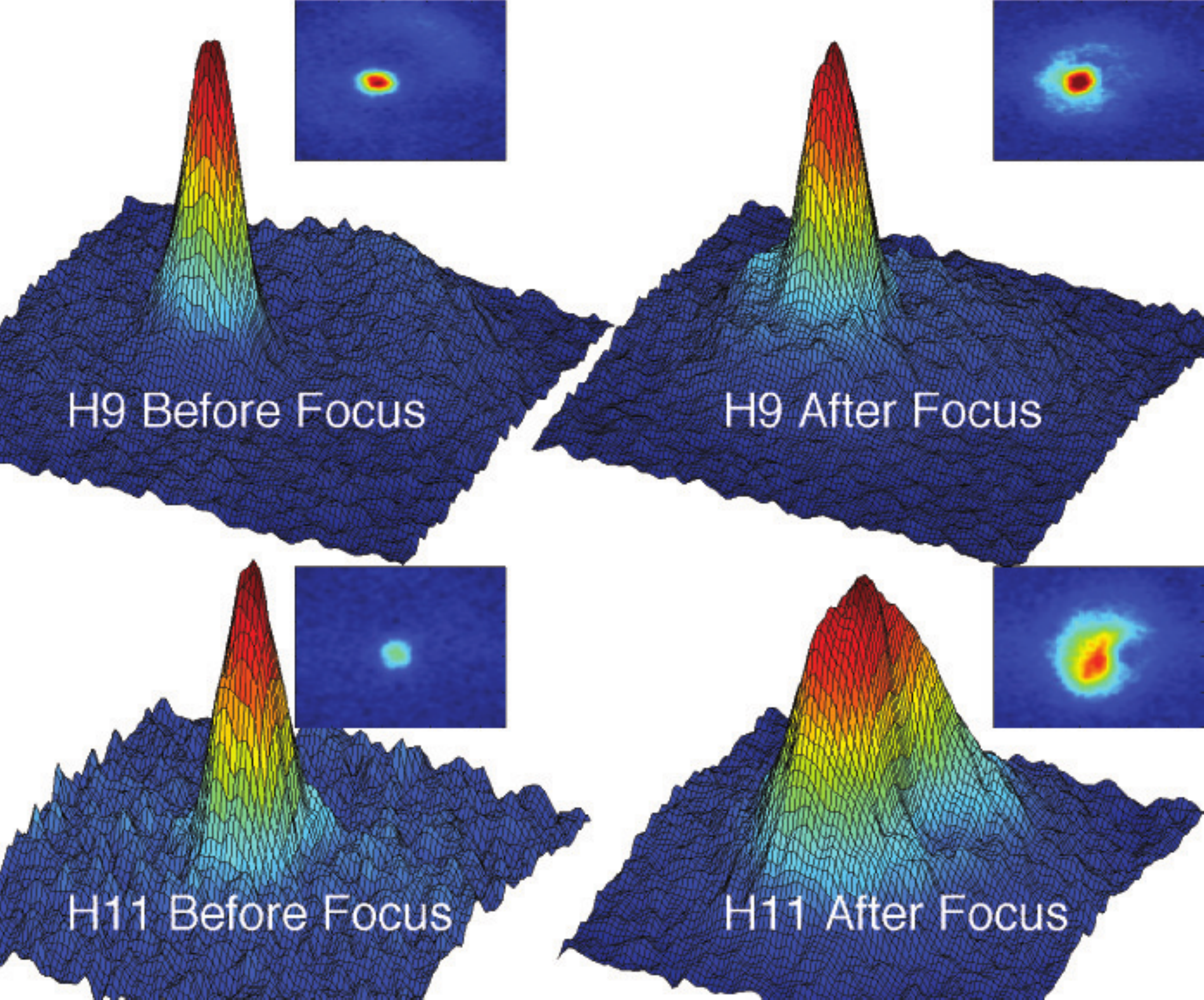}}
\caption{Experimental data for the 9th and 11th harmonics generated in a Ti:Sapphire fsEC XUV source illustrating the XUV beam profiles for gas jet positions before and after the focus, and low xenon pressure (10 Torr in the interaction region).  Variation in phase matching in these regions translates into different far-field spatial patterns, which include effects from long and short trajectory harmonic generation \cite{Hammond2011}.}
\label{fig:HHGBeamShape}       
\end{figure}

The role of multiple quantum pathways is particularly important for the development of XUV frequency combs, where the coherence of the generated light is critical, and the coherence of long trajectory harmonic generation is known to be reduced compared to that of the short trajectories.  The role of coherence has already been discussed in the context of XUV direct frequency comb spectroscopy \cite{Cingoz2012} and will also be important in XUV dual comb \cite{Lee2011} spectroscopy applications.  It is therefore appealing to further develop numerical modeling to study the macroscopic aspects of high harmonic generation. The on-axis simulations of harmonic yield that we have performed show that phase matching affects long and short trajectory harmonic generation differently at low gas pressure (several tens of Torr) and longer interaction lengths (300-500 $\mu$m) at the driving field focus. Moreover, the phase matching is harmonic dependent and the long and short trajectories are favorably phase-matched under different experimental conditions \cite{Hammond2011}.   Extension of this on-axis theory to compute the far-field beam profile will help to connect theoretical predictions with the measurements, such as for the data shown in \fref{fig:HHGBeamShape}. This data illustrates that when the gas jet is positioned before the focus, the long and short trajectories are largely separated, and when positioned after the focus, the trajectories overlap.  This feature may be useful in some XUV frequency comb applications.  Preliminary results of a model including phase matching in three dimensions to compute the far-field beam profiles shows qualitative agreement with the measurements, and they indicate that the relative contributions of the different generation pathways in the experimental observables, such as beam shape and harmonic flux, are very sensitive to the parameters used in a particular experiment.  In this way, the low pressure regime can be utilized to explore the role of separate quantum trajectories in the HHG process for near-threshold harmonics \cite{Yost2009}.

\subsection{The XUV Output Coupler}
\begin{figure}[h!]
\centering
\resizebox{0.75\textwidth}{!}{
 \includegraphics{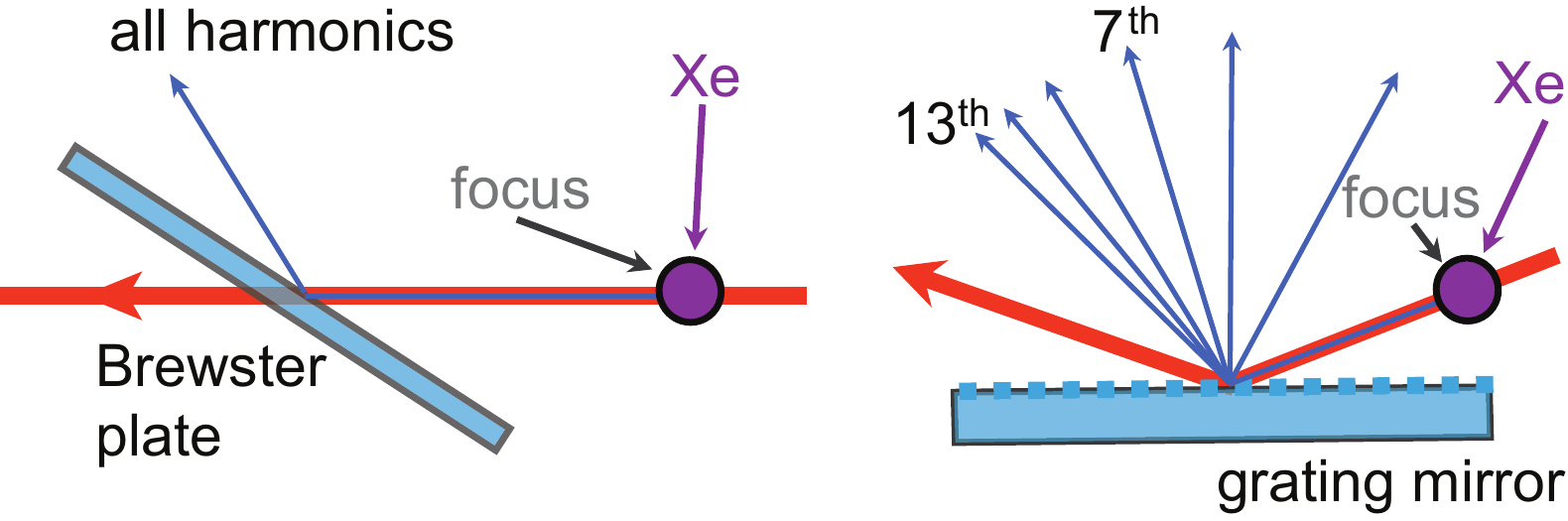}}
\caption{Illustration of Brewster plate (left) and grating mirror (right) XUV output coupler designs. In each case, the gas target is placed near the intracavity focus where the harmonics are generated, and the output coupler is placed a small distance away from the focus.  The out-coupled harmonics are collinear in the Brewster plate output, while they are spectrally dispersed for the case of the grating mirror.}
\label{fig:OutputCouplers}       
\end{figure}

In the HHG process, the XUV is generated colinearly with the optical driving field. Thus, one of the fundamental challenges to fsEC development is the design of a high-efficiency XUV output coupler. The design criteria for the XUV output coupler in fsEC applications are covered comprehensively in \cite{Pupeza2011}. The technical requirements for the XUV output coupler are extremely demanding as it must efficiently couple the XUV out of the cavity while withstanding the high average power and peak intensity in the fsEC, and at the same time it must contribute negligible loss and dispersion as well as low nonlinearities for the circulating cavity field.    The two output couplers that have been used most successfully in fsEC's are the Brewster plate and diffraction grating output coupler, and are illustrated in \fref{fig:OutputCouplers}.

The first demonstrations of fsEC-based XUV generation used Brewster plates to extract the harmonics from the cavity \cite{Jones2005,Gohle2005}.  This method relies on the different refractive indices for the fundamental and harmonic wavelengths.  The plate is aligned at Brewster's angle for the fundamental beam to eliminate Fresnel reflection, but the harmonics have different reflection coefficients and a portion of each harmonic is coupled out of the cavity. Thin Brewster plates are chosen to limit the dispersion added to the cavity, and for a moderate finesse fsEC ($\mathcal{F}<500$) the dispersion can be compensated without the use of chirped mirrors \cite{Lee2011}. Normally, a thin plate (\emph{e.g.} 250 $\mu$m) of sapphire is used, although other appropriate materials also exist \cite{Gohle2006}. The efficiency of the Brewster plate output coupler depends on wavelength and material used for the plate but varies from 5-15\% for the harmonics of interest.  The output-coupled XUV beam in this case consists of many harmonics superimposed on one another.  This situation may be desirable for some applications, while for others a diffraction grating must be used to separate the harmonics which causes additional losses for the XUV. One experimental difficulty with this approach is that over time the sapphire plates suffer irreversible optical damage, leading to excessive losses for the optical driving field, and the plates must be replaced.

For applications requiring spectrally resolved harmonics, an XUV diffraction grating output coupler can be employed.  First reported in \cite{Yost2008}, this method utilizes a small-period (\emph{e.g.} $\lambda_g=$ 420 nm) diffraction grating etched in the outermost layer of the dielectric stack mirror to couple the harmonics out of the cavity without significantly affecting the circulating fundamental field.  The grating structure is considered in the design of the dielectric mirror, and does not reduce the high reflectivity of the fundamental light. An important aspect of the grating design is that the sub-wavelength grating can facilitate coupling between the cavity mode and a slab waveguide mode created by the dielectric stack. Fortunately, the coupling can be eliminated by increasing the depth of the grating without degrading the performance \cite{Yost2011}.   The XUV is coupled out of the fsEC with the diffraction grating with an efficiency ranging from a few percent to 10\%, depending on the harmonic order.  More recently, a blazed diffraction grating has been developed that achieves a diffraction efficiency of about 15-20\% from 35-80 nm \cite{Yang2011}.

Another out-coupling method involves passing the harmonics through a small aperture in one of the focusing mirrors \cite{Moll2006}, however this method has been met with limited success.  This approach takes advantage of the fact that the divergence of the harmonics is less than that of the fundamental beam.  The aperture must be small to mitigate loss and diffraction due to the high finesse requirement for intracavity HHG, making alignment of the harmonic through the hole difficult.  It was proposed that a higher order intracavity mode, such as TEM$_{01}$, could be used so that the fundamental beam has negligible field amplitude along the axis of propagation, and thus could avoid the aperture.  However, due to the increased spot size and increased Gouy phase shift for higher order modes, the HHG efficiency decreases.  Conversely, intracavity mode tailoring has been proposed to remedy these limitations \cite{Weitenberg2011}.  Here, degenerate cavity modes are used, which allow for designing the aperture to be large without causing considerable loss, while the field is at a maximum along the beam axis near the focus.  To achieve mode degeneracy within a non-confocal cavity, the focusing mirror separation is carefully tuned such that the increased phase shift experienced by higher order modes can be compensated, allowing for a degeneracy of several mode orders.

Another intriguing method proposed for out-coupling the harmonics is non-collinear HHG \cite{Moll2006, Wu2007, Ozawa2008a}.  Here, two pulses from the fundamental field interact at the focus at a small angle; the harmonics generated propagate in the direction bisectrix of the fundamental beams.  The difference in the direction of the harmonics and fundamental beams allow for the harmonics to be out-coupled while minimally affecting the cavity finesse. Due to practical issues regarding the timing of colliding pulses, among others, this technique has yet to be experimentally demonstrated. 

With better understanding of the limitations of the diffraction grating and Brewster plate as XUV output couplers, new beam separators have been developed.  One method, in the vein of the diffraction output coupler, has a small angle top layer wedge on a highly reflecting mirror \cite{Pupeza2011}.  Because of the small angle, the wedge has minimal effect on the reflectivity of the mirror beneath for the fundamental beam, whereas the XUV beam is reflected by the wedge layer, allowing for a small divergence of the fundamental and harmonic beams.  Although this output coupler has shown to have similar scalability and coupling efficiency to the diffraction grating, the harmonics are co-linear, which may have benefits for some applications.  Another design uses the large Fresnel reflectivity at grazing angle incidence on an intracavity optic to increase the out-coupling to greater than 50\% for XUV light down to 15 nm \cite{Pronin2011}. Here, the incident angle is increased to $ >$75$^\circ$ to increase the XUV reflectivity, where the front and back of the plate have an anti-reflection coating for the fundamental beam at the grazing angle.  The high reflectivity makes this a promising method of coupling out the XUV light, although the limitations of this design may be comparable to those of the Brewster plate (that is, dispersion and heating/optical damage for high power scalability). In addition, these XUV output coupler designs will also be of use for lower repetition rate, traditional CPA single-pass HHG XUV systems that currently use gratings and aluminum filters to separate harmonics from the fundamental driving field as well as from each other. Indeed, for most applications of HHG generated XUV radiation, both of these filtering steps must be made.

\section{Experimental Details}

\subsection{fs seeds for fsEC}
The first demonstrations of XUV generation via fsEC employed Ti:Sapphire mode-locked oscillators  to seed the cavity \cite{Jones2005, Gohle2005} and resulted in intracavity powers up to 480 W and pulse energies of 4.8 $\mu$J. As HHG is highly nonlinear there is a large payoff for increasing these performance levels. An early effort to increase the intracavity pulse energies was made by decreasing the fsEC cavity length and seeding it with a chirped pulse oscillator operating at 10.8 MHz. While some success was reported \cite{Ozawa2008}, the XUV power did not substantially improve. Moreover, in our experience decreasing the repetition rate much below 50 MHz leads to difficulties locking the cavity length due to a decrease in the width of the fsEC's resonances and increased mechanical instability of such long cavities. In addition, these steps can be counterproductive for XUV frequency comb applications as  high power per comb element with adequate spacing of the elements is desired. A second approach to increasing intracavity intensities is simply to increase the average power of the input seed by using either a CPA fiber-based system operating at its full repetition rate (100-150 MHz) or injection-locking a solid-state mode-locked oscillator to a secondary gain cavity. These latter two approaches are discussed in more detail in \sref{sec:fsEC_results}.  As a final note, while increasing the fsEC finesse is also capable (in theory) of boosting the intracavity intensity, it subsequently leads to experimental difficulties such as troubles in cavity length locking as well as  higher sensitivity to intracavity dispersion (discussed in \sref{sec:Dispersion}) and mirror damage (see \sref{subsection:MirrorDamage}). For these reasons, a finesse of approximately 500 is now typically used and intracavity intensities are obtained by using high power seeds as discussed above.

\subsection{Vacuum Apparatus}
A system producing  XUV light must operate in a vacuum environment to avoid absorption of the light by atmospheric gases. While this might seem to suggest that only a rough vacuum is necessary, the low-loss demands of the optical coatings in the fsEC can result in sensitivity to hydrocarbons and other contaminants in the vacuum system and therefore high or ultra-high vacuum (UHV) is a desirable goal to minimize contaminants and degradation of the fsEC mirrors.  Though not fully understood, degradation of the fsEC mirrors are most likely the result of optical and XUV photo-assisted surface hydrocarbon chemistry on the dielectric mirrors. We discuss these aspects in more detail in \sref{subsection:MirrorDamage}.  As a result of this mirror damage we attempt to use only high-vacuum compatible components, and we thoroughly clean all components before introducing them to the vacuum system, including plasma cleaning of the fsEC optics. The first generation system built in our lab is a large, homemade aluminum vacuum chamber with o-ring seals that is pumped with a 500 $L/s$  turbo pump.  Next generation systems will use UHV compatible components and chamber designs to reduce contaminants as well as to provide better pumping of the residual gas used for the HHG process.  We discuss technical details of this aspect in \sref{sec:GasTarget}.

\subsection{Active Stabilization of the Enhancement Cavity}
\label{fsEC:Stabilization}
Active stabilization of CW optical sources to optical reference cavities for frequency metrology and build-up cavities for intracavity harmonic generation has continuously improved over the past three decades \cite{Hansch1980,Drever1983,Notcutt2005}. Beyond the differences that arise in the character of the fsEC resonances, as described in \sref{subsec:fsOptRes}, the principles of stabilizing a mode-locked oscillator to a fsEC are much the same as stabilizing a CW laser to a high finesse optical cavity: one can use a modulation-based method of generating an error signal, such as the Pound-Drever-Hall (PDH) method  \cite{Drever1983}, or one can use a `DC' approach such as the H\"{a}nsch-Couillaud method \cite{Hansch1980}. As discussed earlier in \sref{subsec:fsOptRes}, when exciting a fsEC there are two degrees of freedom that need to be stabilized. An appropriate, although perhaps not perfectly orthogonal, error signal for both the cavity length and the relative carrier-envelope-offset frequency can be derived from the same cavity reflection signal by spectrally resolving this reflection as discussed below.

\subsubsection{Stabilization of Cavity Length}
To stabilize our fsEC such that it is length-matched to the oscillator at the central fringe shown in \fref{fig:ResonanceMap}, we use a modified PDH method, in which a high frequency ($\sim1$ MHz) modulation signal is applied to the fsECs PZT. In practice the exact modulation frequency is varied until a PZT resonance is found which significantly enhances the signal to noise of the error signal. An excellent tutorial to the classic PDH technique can be found in \cite{Black2001}, which includes many technical details as well as fundamental, shot-noise limitations of the technique. It has become common practice to use the PDH name for other modulation-based locking techniques that share similar benefits, such as the method we describe here.  In our case, we do introduce modulation sidebands onto our intracavity intensity, but they are extremely small and are inconsequential.  The error signal is derived via detection of the amplitude modulation on the cavity reflection of the seed laser.  Rather than using the full spectrum of the cavity reflection, a diffraction grating and a narrow slit is used to select a limited spectral bandwidth at the pulse's spectral centre. Adjustment of the exact centre frequency and bandwidth can significantly improve the lock performance, particularly in the case of high finesse fsECs, where insufficient control loop bandwidth results in some residual instability. After the slit, the filtered light is detected with a photodetector with several MHz of bandwidth and subsequently demodulated with a double-balanced mixer to produce the error signal. When the phase of the mixer input signals are adjusted to be in quadrature, a suitable error signal is produced at the output of the mixer and its character is nearly identical to the classic PDH case.

In our case, a small mirror in the fsEC is fixed to a PZT element mounted on a lead-filled copper slug in a similar manner to that described in \cite{Briles2010}. With this configuration we are able to achieve a bandwidth of $>$100 kHz, depending on the cavity finesse. We also employ a slow PZT with a larger dynamic range to compensate for slow drifts due to temperature fluctuations. Although we lock the cavity length of the fsEC to the oscillator, it usually works equally well to lock the oscillator's cavity length to the fsEC. In the latter case one possible subtlety that can be of concern is the cross-coupling of the oscillator's cavity length actuator to also change its carrier-envelope offset frequency \cite{Holman2003}.

\subsubsection{Stabilization of the Relative Carrier-envelope-offset Frequency}
With the cavity length servo engaged, residual fluctuations of $\Delta \fceo$ between the cavity and oscillator can contribute to amplitude modulation on the intracavity power.  As with cavity length stabilization, it is possible to derive an error signal to address this instability by detecting a portion of the spectrally resolved the cavity reflection signal. In this case the error signal is formed by detecting two narrow bandwidth signals located at the spectral extremes of the cavity reflection signal and subsequently subtracting them \cite{Jones2001,JasonJones2004}. An easier, though somewhat compromised (non-orthogonal) simplification of this technique is to derive an error signal for the cavity length lock in one spectral region and derive a second error signal from a different portion of the cavity reflection spectrum for the $\Delta \fceo$ lock.  This latter technique is useful in our high finesse cavity system at 800 nm.  On the other hand, the high power Yb-doped fiber laser systems tend to be sufficiently stable that active stabilization of $\Delta\fceo$ is not necessary, particularly for lower finesse fsECs.  The actuators used to stabilize $\fceo$ are the same as for other femtosecond frequency comb applications, including control of the oscillator's pump power or by shifting the entire frequency comb externally via an acousto-optic modulator.

\subsection{Gas Target for HHG}
\label{sec:GasTarget}
In \sref{Theory:PhaseMatching} we reviewed the theory behind generation of harmonics in the tight-focus regime, which is required in a fsEC to achieve the intensity required for HHG ($>$10$^{13}$~W/cm$^2$).  In this regime efficient generation of harmonics requires the gas target to be restricted to a small volume near the focus of the fundamental driving field in the fsEC.  In this section, we review two nozzle geometries for the HHG gas target that have been used in previous demonstrations of fsEC-based HHG \cite{Jones2005,Gohle2005,Ozawa2008} (both are routinely employed in CPA-based HHG systems), and a third geometry that is a hybrid of the two nozzle geometries. We also discuss the results of computational fluid dynamics (CFD) simulations for these nozzle geometries that help with the evaluation and optimization of these nozzle designs for use in fsEC HHG systems.

\begin{figure}[h!]
\centering
\resizebox{0.8\textwidth}{!}{
  \includegraphics{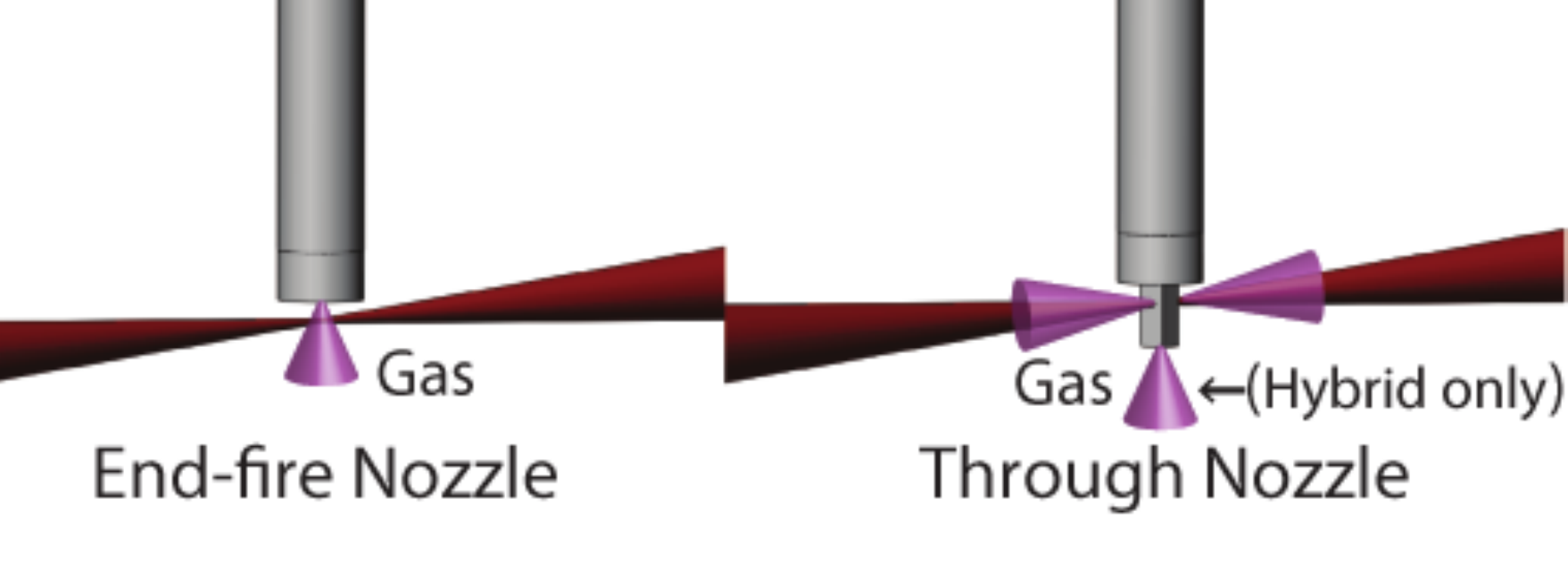}}
\caption{Illustration of gas nozzle geometries with the end-fire nozzle (left) and the through-nozzle (right) designs shown with the laser beam (red) focused near the gas nozzle.    The hybrid-nozzle design is a modified through-nozzle in which the end of the nozzle is opened to allow gas flow to escape into the vacuum in order to increase the transverse velocity of the gas through the interaction region.}
\label{fig:EndFire}       
\end{figure}

In a previous demonstration of intracavity XUV generation, the authors of  \cite{Ozawa2008} used a \emph{through-nozzle} geometry, in which the laser field is threaded through a small hole drilled through a stainless steel tube containing xenon gas.  In the second geometry \cite{Jones2005,Gohle2005}, a nozzle is formed by a small hole in a gas line and the xenon gas is allowed to freely expand into vacuum.  This \emph{end-fire nozzle} is placed as closely as possible to the intracavity beam in order to maximize the density of gas interacting with the laser field.  The geometry of these two gas nozzles is illustrated in \fref{fig:EndFire}.

Each of the designs has attractive features as well as drawbacks.  The end-fire nozzle is technically less challenging to implement, but  the nozzle must be located sufficiently far from the intracavity focus so as not to introduce excessive loss in the cavity.  Much higher gas density is possible for the through-nozzle, but the nozzle must be precisely positioned with respect to the intracavity field, and the apertures that the laser is threaded through must be large enough not to introduce significant loss.  Furthermore, the apertures of the through-nozzle allow gas to exit the nozzle along the optical axis, which increases amount of plasma generated and increases the absorption of XUV by the neutral gas atoms. Both of these consequences are undesirable, but the excess plasma is a major problem for intracavity HHG, as we will discuss in \sref{sec:PlasmaEffects}.

In order to better understand how the geometry of these different nozzle designs affect the generation of harmonics we performed CFD simulations to determine the gas density distribution for a variety of nozzle parameters, which is used in the computation of the harmonics reviewed in \sref{Theory:PhaseMatching}.  We presented a detailed summary of these results for the end-fire nozzle in \cite{Hammond2011}.  We have also performed a similar analysis for the through-nozzle geometry, and a third geometry or \emph{hybrid-nozzle} that eliminates the stagnant flow at the center of the through-nozzle by opening a hole at the end of the tube that allows gas to flow from the end as it does in the end-fire geometry \cite{Hammond2011a}.  We do not give a comprehensive summary of the simulations here, but we do summarize the properties of each of the nozzles in \tref{Table:GasNozzle} in terms of gas density, interaction length, and the transit time or interaction time.

\begin{table}[hbt]
\caption{\label{Table:GasNozzle}Summary of the properties of three different nozzle geometries with respect to the gas density in the interaction region, effective gas-laser interaction length, and the transit time of atoms through the interaction region due to gas flow along or across the intracavity laser field.}
\begin{indented}
\item[]\begin{tabular}{|c|c|c|}
\hline Nozzle Geometry & Advantages & Disadvantages \\
\hline & & Large interaction length \\
 Through-hole & High density & Long interaction time \\
 & & Low gas velocity \\
\hline Hybrid& High density & Large interaction length\\
  &  & Long interaction time  \\
\hline  & Small interaction length & Density falls rapidly \\
 End-fire& Shortest interaction time &  with distance from orifice \\ 
  & No gas flow along laser axis & \\
\hline
\end{tabular} 
\end{indented}
\end{table}

\begin{figure}[h!]
\centering
\resizebox{0.75\textwidth}{!}{%
  \includegraphics{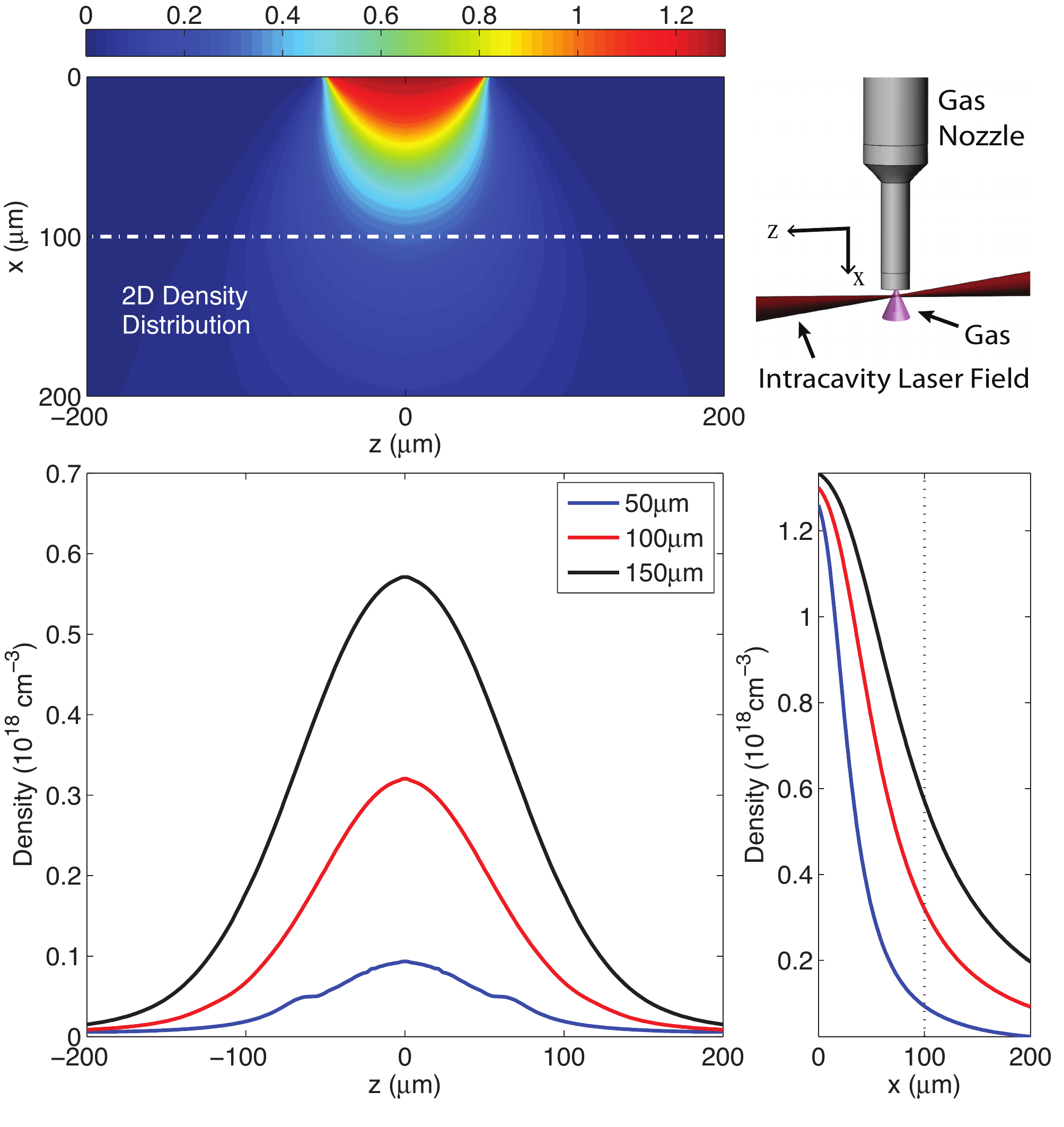}}
\caption{CFD simulations of the spatial dependence of gas density ($\rho$) for the end-fire nozzle with diameters of 50, 100 and 150 $\mu$m, for a capillary tube-like design.  The upper left panel shows the 2D density distribution of the gas emerging from a 100 $\mu$m nozzle, $\rho(x,z)$.  The white line indicates the x-location for which the z-dependence of the density is plotted. The color bar is in units of density (10$^{18}$ cm$^{-3}$).  The lower left panel displays $\rho(z)$ along the optical axis (z-axis), for each nozzle, at a vertical distance of 100 $\mu$m from the nozzle tip. The lower right panel displays the decrease in $\rho(x)$ with vertical distance from the nozzle opening. The gas nozzle is shown relative to the intracavity laser field in the upper left panel.}
\label{fig:GasDensity}       
\end{figure}

In \fref{fig:GasDensity} we present a series of CFD simulation results for the gas density distribution for three end-fire nozzles with 50~$\mu$m, 100~$\mu$m, and 150~$\mu$m diameters. The geometry simulated in this figure is identical to a capillary tube style nozzle, or simply a long cylindrical tube. This geometry is slightly modified from the geometries simulated in \cite{Hammond2011}, but the results are only slightly different for the nozzle diameters of interest here.  In the figure, the top left panel shows the 2D density distribution of gas, the lower left panel shows the variation in density along the laser propagation axis at a distance of 100~$\mu$m from the nozzle tip, the lower right panel shows the gas density drop as a function of distance from the tip of the nozzle, and the upper right panel shows the geometry of the gas nozzle relative to the intracavity laser field. Practically speaking, in a fsEC the nozzle can only be placed as close as a few tens of $\mu$m from the intracavity focus to avoid diffraction losses that spoil the cavity finesse.

In our experience, the end-fire nozzle is the most suitable choice of nozzles for fsEC due to the elimination of gas flow along the optical axis and the decreased interaction time between the laser and the gas atoms.  We have worked experimentally with the through nozzle geometry, and we found it difficult to reproducibly deliver XUV beams with high quality spatial modes from the system, and plasma-induced oscillations were a serious limitation.  The hybrid nozzle we tested represents a slight improvement over the through nozzle, but about 10\% of the gas still flows along the laser propagation direction and leads to significant plasma oscillation.  The end-fire nozzle, on the other hand, reliably produces high quality beams, and while the harmonic generation is extremely sensitive to alignment of the nozzle, it is much more straightforward to align. Interestingly, very recent work using a 100 kHz single-pass Ti:Sapphire CPA system in the tight-focus regime has measured the XUV flux for the 15th and 17th harmonics for the through-nozzle and end-fire nozzles \cite{Heyl2012}. Their results agree with the observations we have reported for the fsEC case, but the disadvantages of the through-nozzle geometry are not as severe in the single-pass geometry as they are in the fsEC. When singe-pass repetition rates beyond a few MHz are used, we expect the through-nozzle to become even more disadvantageous in this case.

We are currently using thin-walled quartz capillaries for our end-fire nozzles, with an inside diameter of 100-150 $\mu$m.  Alignment of even smaller nozzles becomes very challenging, but a more fundamental limitation is that for smaller nozzle diameters, the gas density drops extremely quickly and the expansion of the gas begins to limit how small the interaction length can be in the laser field, some tens of micrometers away from the nozzle orifice \cite{Hammond2011}.  Alignment of the 150 $\mu$m nozzles is simplified with the small outer diameter (250 $\mu$m) compared to thick walled capillaries, or metal tubes of the same inside diameter.  Similar gas nozzle geometries are also used in recent high XUV flux fsEC systems, where a 100 $\mu$m and 300 $\mu$m stainless steel nozzle \cite{Lee2011}, and 100 $\mu$m glass nozzle \cite{Yost2011} have been used.

Finally, we note that excess gas that accumulates in the vacuum system due to imperfect pumping can cause a number of problems, but most troublesome of all is the reabsorption of the harmonics.  Different methods have been employed to expel this gas, but both involve catching the gas before it can diffuse into the entire volume of the vacuum chamber.  In the work of \cite{Lee2011} a turbo pump is placed in the base of the vacuum system immediately below the gas nozzle, while in the work of \cite{Yost2011}, a special gas catch assembly is placed a small distance below the gas nozzle and connected to a vacuum roughing line \cite{Yost2011a}.  We have implemented a similar \emph{gas catch} arrangement in our systems, and we have observed significantly improved XUV flux when it is properly aligned.

\subsection{Detection of Harmonics}
\begin{figure}[h!]
\centering
\resizebox{0.75\textwidth}{!}{%
  \includegraphics{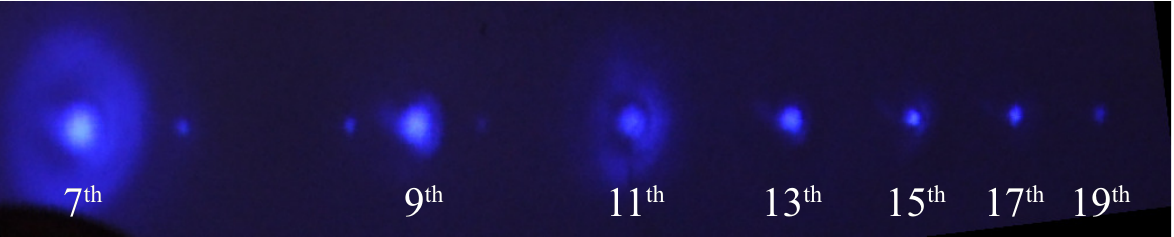}}
\caption{Image of sodium salicylate phosphor illuminated by harmonics 7-19 from a Yb-doped fiber laser fsEC operating near 1040 nm. The spots between the 7th and 9th harmonics are higher diffraction orders.}
\label{fig:YbBlueDots}       
\end{figure}

Detecting the harmonics generated by the fsEC source can be challenging for the lower order harmonics, where detectors and filters must be able to detect the low harmonic power without detecting the on-axis scattered fundamental light.  For wavelengths shorter than 80 nm, aluminum filters directly deposited on silicon photodetectors work very well, but we have not the same level of success filtering out the 800 nm fundamental light with directly deposited filters for the longer wavelength harmonics.  To solve this problem, photomultiplier tubes with phosphor-coated windows can be used, and it is also common to use a phosphor screen to image the fluorescence to obtain beam profiles for the harmonics.  The linearity and spectrally flat responsivity of phosphors such as sodium salicylate makes this method convenient for comparing relative power from the harmonics as well as measuring the beam profile of the harmonics.  An example image is shown in \fref{fig:YbBlueDots}.

There is currently great interest in developing new detectors for  XUV light.  We are using our fsEC systems to evaluate the performance of diamond and selenium photoconductors.  Diamond detectors are particularly interesting for detection of high energy photons for a variety of reasons, including their radiation hardness, low leakage current, and visible light insensitivity (solar-blindness) \cite{Pace2006}.  This last factor is of particular interest for fsEC-based XUV generation where the scattered light from the fundamental near-infrared light in the cavity can be orders of magnitude brighter than the generated harmonics.  Selenium is also used in digital x-ray radiography as direct, electronically readable x-ray detectors \cite{Chotas1999}.  Development of selenium detectors for enhanced short wavelength visible radiation detectors with extremely low response in the near infrared is also progressing \cite{Wang2010}.  This recent success has sparked interest in evaluating the use of these photoconductors in the VUV and XUV spectral regions.  Preliminary results with the 5th harmonic (158 nm) are promising, and demonstrate excellent extinction of scattered fundamental light at 800 nm.

\section{Experimental Challenges with fsECs}
\label{sec:fsEC_ Challenges}
In the previous section we described the experimental details of individual components of the apparatus and generation of harmonics in the tight focus of the fsEC.  In this section we discuss the challenges that arise when the individual components are integrated and begin to interact with one another when high power is present in the fsEC. Several factors have been identified as essential to increasing the XUV photon flux and photon energies. The factors described in this section can be classified into one of two primary areas, namely 1) the ability to maintain the high reflectivity of fsEC mirrors under high peak intensity and high average power for extended periods of time, and 2) managing the various effects that arise in the fsEC when a plasma is present, including stabilization of the fsEC and generation of harmonics via HHG at extremely high repetition rates.

\subsection{Mirror degradation and damage}
\label{subsection:MirrorDamage}
As the mirrors within an fsEC must be highly reflective, even a slight degradation in their performance can be of concern. For the most part, at the intracavity average power and intensity achieved in fsEC systems the mirror reflectivity is often degraded rather than irreversibly damaged. In high power Yb-doped fiber systems mirror damage is reported for high average power (18 kW) and pulse durations shorter than 640 fs \cite{Pupeza2010a}.  The damage is not described in detail, but rather a subject for further investigation.  In another Yb-doped fiber laser system \cite{Yost2011} mirror damage is not a limiting factor for intracavity average power of $>$7 kW, and the mirror degradation can be reversed with an oxygen plasma \cite{Yost2011a}.  The high power Ti:Sapphire system reported in \cite{Lee2011} has issues with the permanent damage of the Brewster-plate XUV output coupler, which is located relatively close to the intracavity focus and is subject to much higher intensities than other optics in the cavity.  

Optically induced degradation of the mirrors can arise via two sources. The first results from high (several kilowatt) intracavity average power and very high peak intensities of the optical driving fields on mirror surfaces. A second is degradation induced by the relatively high photon flux XUV generated in the fsEC.  From work on synchrotrons and other advanced light sources, it is well-known that XUV and x-ray optics suffer degradation by contaminant hydrocarbons in the vacuum that deposit on mirror surfaces and interact with XUV high energy photons (see, for example, \cite{Pereira2009} and references therein).  Cracking of the hydrocarbons and attraction of new hydrocarbons to the surface from the surrounding vacuum leads to run-away build-up on and reduced reflectivity of the mirror surface. There is speculation that dangling bonds from the specific dielectric materials also play a role in the process. These effects can often be reversed with the introduction of oxygen gas or ozone near the mirror surface.

Such processes have been widely investigated, but recent work \cite{Pereira2009} that investigates the surface contamination of silica by toluene (C$_7$H$_8$), in the presence of 213 nm UV light is particularly reminiscent of the effects we observe in our fsEC systems  (\fref{fig:Contamination_Revival} below).  The measurement relies on the detection of photoluminescence (PL) of the silica produced by UV light, and the decay of the PL signal via absorption within the layer of hydrocarbon contaminants as they accumulate on the silica surface.  After a large reduction in PL signal is observed, the toluene gas is turned off and oxygen is introduced.  In the presence of the 213 nm radiation, photolysis of the oxygen occurs and a photochemical reaction results in the formation of ozone (O$_3$), which etches the toluene-based photo-deposits and results in a revival of the PL signal.  It would appear that the effects observed in fsEC systems is a multi-photon analogy of these experiments. This is not at all unreasonable given the high intensities present in the fsEC and bond energies involved in the hydrocarbon dissociation.  The C-H and C-C bonds have dissociation energies of about 4.3 and 3.6 eV, respectively \cite{Pereira2009}, each of which would the energy of three photons in the 800 nm Ti:Sapphire system, or a two-photons for the 520 nm fsEC system described below.

From our experience a similar hydrocarbon based degradation process occurs in fsECs and can be mediated by both optical and XUV photons. Focusing first on degradation induced by the optical driving field, we have observed varying degrees of degradation in our fsEC systems operating at 800 nm and 1040 nm depending on the particular mirrors used (from different vendors) and the cleanliness of the vacuum environment. In general, we find the problem is much more pronounced at 800 nm.  In our Ti:Sapphire laser seeded system operating near 800 nm, different mirror coatings (\emph{i.e.} different vendors) have different degradation characteristics, even when no intracavity plasma (and therefore no XUV light) is generated.  Mirrors from one vendor exhibited a lifetime of several minutes to a few tens of minutes, depending on the intracavity intensity.  We show an example of this degradation in \fref{fig:Contamination_Revival} for mirrors from this vendor.  Mirrors from another vendor have shown lifetimes in excess of several hours at the highest intracavity average power we can achieve with this particular mirror set ($\sim 300$ W).

Once XUV light is generated, the mirrors from this second vendor degrade from their optimum performance in a matter of minutes. XUV degradation is usually limited to the one or two mirrors following the gas target. It remains to be seen to what extent vacuum cleanliness can resolve these issues. Our Yb-doped fiber laser system operating near 1040 nm exhibits substantially slower degradation.  The system shows signs of mirror degradation when the system has been operating at an average power of 2-3 kW for several hours. This system also maintains its performance levels for substantially longer periods of time during XUV production than our 800 nm system.

\begin{figure}[h!]
\centering
\resizebox{0.75\textwidth}{!}{%
\includegraphics{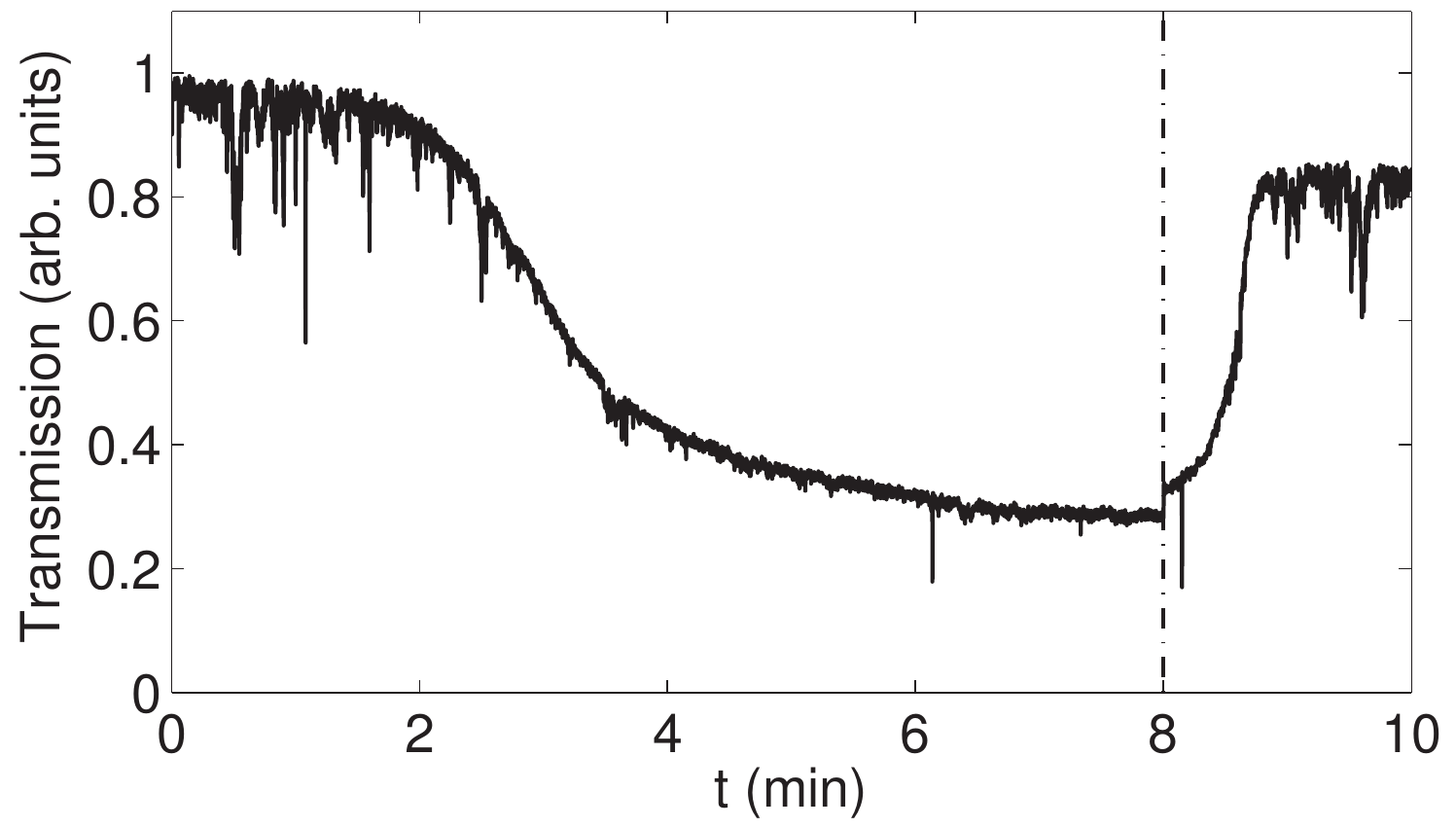}}
\caption{Data showing the reduction in cavity transmission as the cavity mirrors degrade due to hydrocarbon contamination in a Ti:Sapphire system operating near 800 nm with no XUV generation, for a mirrors from a particular vendor.  At $t=8$ minutes (indicated with a vertical dotted line), 6 Torr of oxygen is introduced into the vacuum system and after the cavity lock is engaged, the mirror reflection is restored.  The cavity enhancement is restored in under one minute, and for this data set, it recovers to about 90\% of the value before the mirrors are degraded, although the maximum transmission seen in the figure is about 85\%, limited by the dispersion of the oxygen in the cavity.}
\label{fig:Contamination_Revival}       
\end{figure}

Similar to other XUV and x-ray optics, once degraded the fsEC mirrors can be rejuvenated by pressurizing the cavity with oxygen. This healing process can be significantly accelerated if the cavity is locked such that there is high intracavity optical intensity along with the oxygen atmosphere. The data shown in \fref{fig:Contamination_Revival} shows this revival, where 6 Torr of oxygen is introduced to the vacuum system at $t=8$ minutes.  We have not observed such dramatic degradation or revival with our 1040 nm system, but weekly removal of the mirrors from the vacuum and cleaning with an oxygen plasma cleaner is helpful. Interestingly, a fsEC operating near 520 nm was recently reported \cite{Bernhardt2012} that must be operated in an oxygen environment to avoid mirror degradation. Without the oxygen the cavity enhancement decreases exponentially with a time constant of about 20 seconds. In addition to the obvious change in photon energy for these three cases (1040 nm, 800 nm and 520 nm), the specific dielectric materials and growth techniques used in the mirror construction at different wavelengths will also play a significant role in determining the mirror sensitivity to hydrocarbon contamination. Unfortunately, further quantification of the degradation mechanism from a material perspective is difficult given the unwillingness of mirror vendors to provide their mirror design specifications.

\subsection{Plasma effects}
\label{sec:PlasmaEffects}
The plasma generated during the HHG process plays an important role in the phase matching process and must be managed to generate harmonics most efficiently.  For kHz repetition rate CPA systems the harmonic generation becomes limited when the fractional ionization that occurs during a single femtosecond pulse exceeds a certain level, but issues related to the accumulation of plasma resulting from multiple pulses do not arise because the long time between pulses.  Such is not the case for fsEC systems where the pulse period ($\sim10$ ns) is shorter than the timescales relevant to plasma decay, and even the transit time for the room temperature gas atoms traversing typical focal volumes in the fsEC may exceed 10 pulse periods.

\begin{figure}[h!]
\centering
\resizebox{0.75\textwidth}{!}{%
 \includegraphics{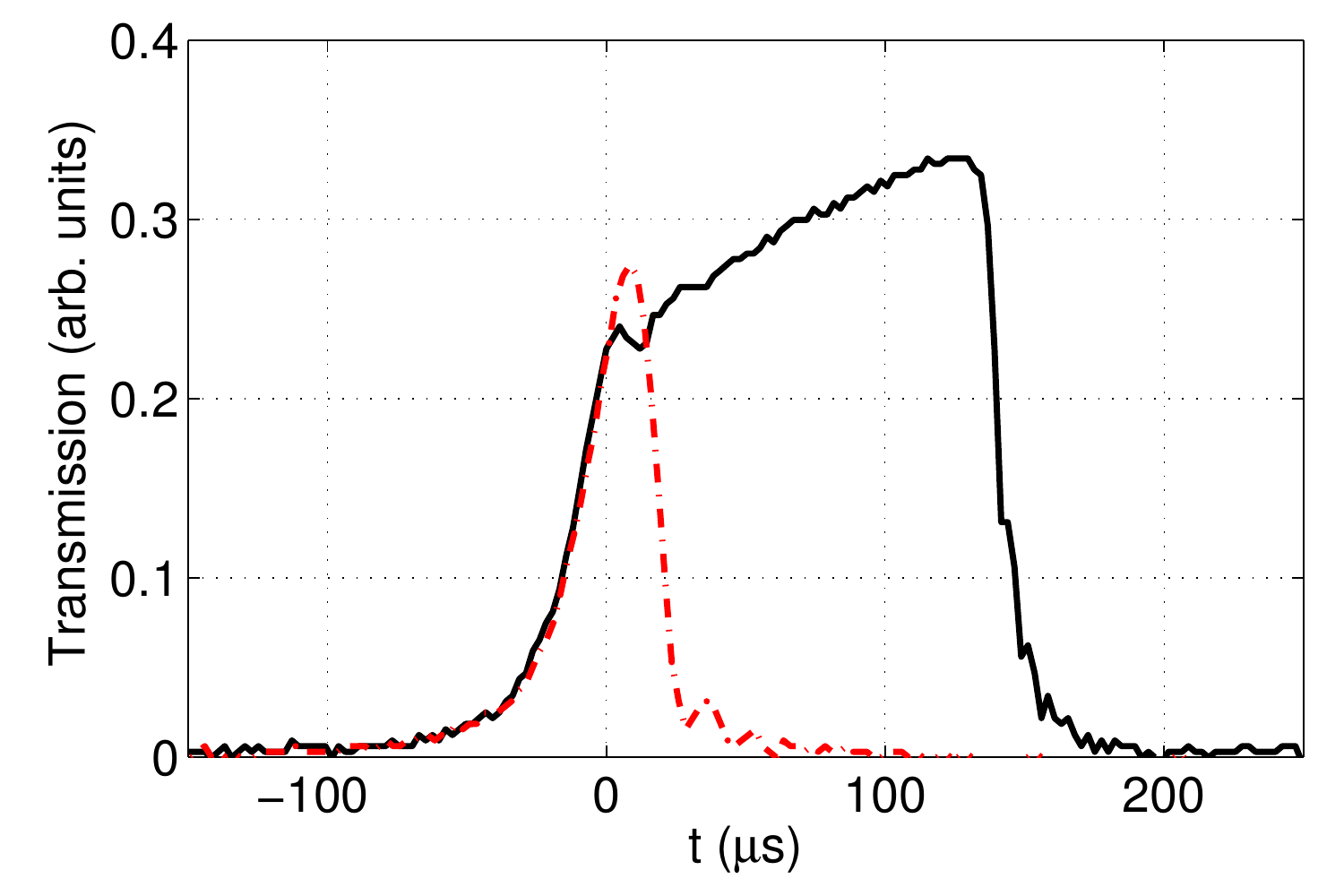}}
\caption{Resonance lineshapes of the transmission from the fsEC when the cavity length is scanned across resonance in the presence of xenon gas for the through-nozzle geometry.  For the solid curve, the change in cavity length compensates for the plasma phase shift as the plasma accumulates. For the dot-dashed curve the cavity length scan is in the opposite direction, which does not negate the phase shift due to the growing plasma, and the cavity rapidly shifts away from resonance.}
\label{fig:PlasmaScan} 
\end{figure}

The generation of intracavity plasma and the limitations it imposes on XUV generation has been studied recently \cite{Carlson2011,Allison2011}.  Over short timescales (\emph{i.e.} the duration of a pulse), nonlinear phase shifts (\emph{e.g.} self-phase modulation) limit the peak intensity of the pulse circulating in the fsEC, and over longer timescales (tens of pulse periods and longer) the steady state plasma density also contributes a phase shift that significantly shifts the cavity resonance.  There is little that can be done about the phase shift resulting from the dynamically generated plasma, and this factor may be one of the fundamental limits of fsEC-based XUV generation. The phase shift due to the steady state plasma can shift the fsEC resonance well in excess of the cavity linewidth.  To demonstrate the effect of the plasma phase shift, we present a plot of the cavity lineshapes when the cavity length is scanned in opposing directions, in \fref{fig:PlasmaScan}. For the solid curve, the change in cavity length partially compensates the increasing phase shift due to the plasma, and the cavity remains on resonance longer than it would without the plasma present.  For the dot-dashed curve, the cavity scan is in the opposite direction, which adds with the plasma phase shift and causes the cavity to move away from resonance more rapidly. The servo control loop can compensate for the phase shift due to the steady state plasma, but even small perturbations in the intracavity intensity can cause a drop in the plasma density that shifts the resonance condition too rapidly for the control loop to follow.  This condition initiates a bistable oscillation in which the servo loop acts together with the plasma build-up and decay to drive the system into and out of resonance.  The mechanism for this oscillation was described in \cite{Yost2011,Allison2011}, where it was also pointed out that locking slightly away from the peak of the cavity resonance improves the stability of the system and can prevent this bistable oscillation.

\begin{figure}[h!]
\centering
\resizebox{0.75\textwidth}{!}{
\includegraphics{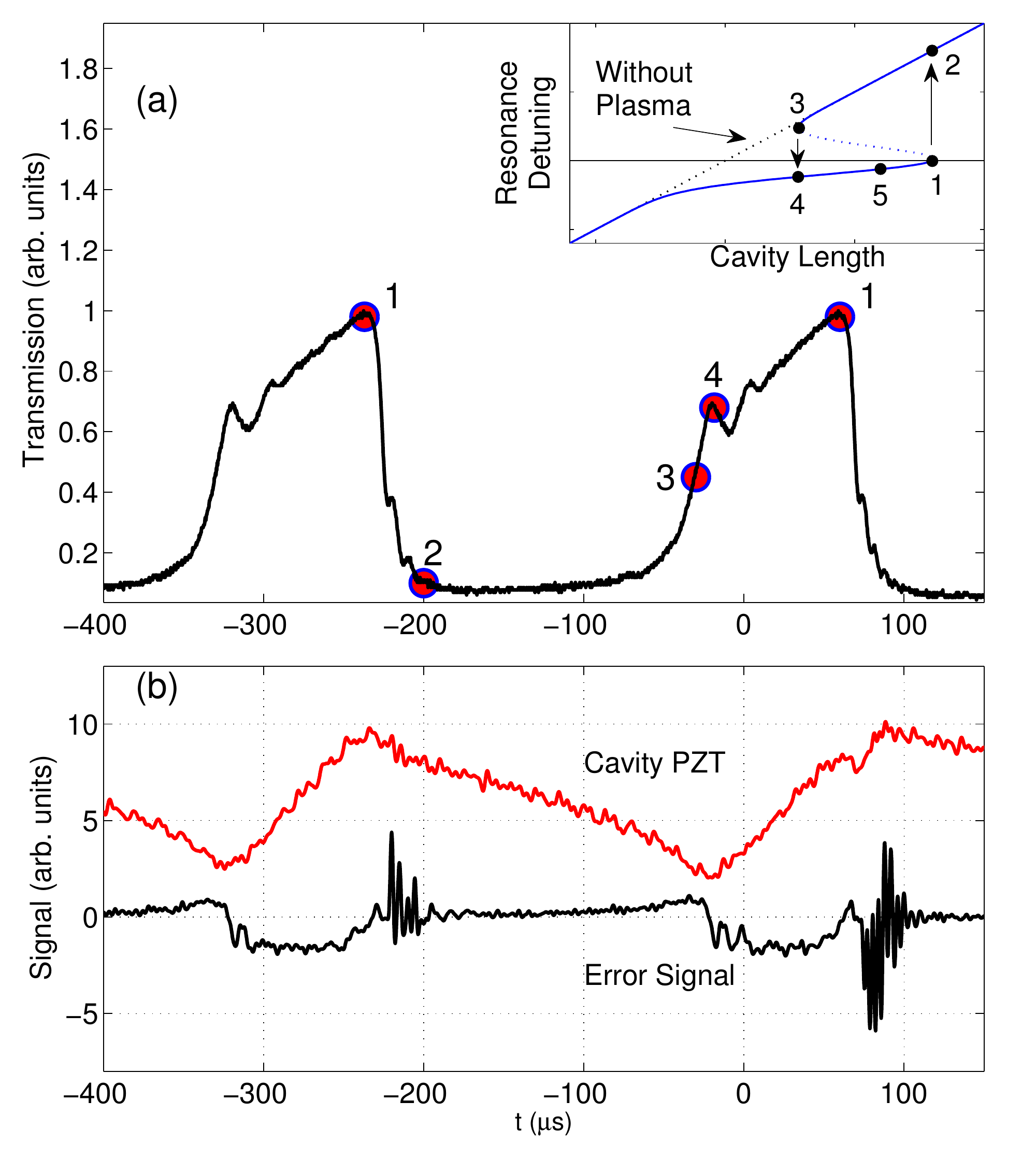}}
\caption{a) Cavity transmission, b) Cavity PZT drive signal and control loop error signal while the servo control loop undergoes oscillation due to plasma induced optical bistability.  Inset: diagram of the bistable loop oscillations that can occur in a fsEC. See text for details of the bistable oscillation.  Point 5 indicates the locking point where stable locking can occur.}
\label{fig:PlasmaOsc}   
\end{figure}

We show an example of large plasma oscillations in \fref{fig:PlasmaOsc}, observed in our Ti:Sapphire seeded fsEC system. When the cavity is locked to the transmission maximum, the on-resonance cavity length differs significantly from its value without plasma, due to the phase shift from the plasma. Because the ionization rate is a highly nonlinear function of intensity, even small perturbations of the intracavity intensity can cause a large and rapid decrease in plasma density, and likewise a rapid shift in the resonance detuning.  Such an event can drive the system into the following oscillation sequence.  Starting from the point of maximum cavity transmission (labelled point 1 in the inset of \fref{fig:PlasmaOsc}), the cavity is perturbed from resonance, which causes the plasma to disappear and the cavity resonance to shift moving the system to point 2.  The servo begins to move the cavity length closer to resonance, corresponding in the figure to a decreasing cavity PZT signal.  As the cavity length approaches resonance, the error signal is observed to acquire a slightly more positive signal and the cavity transmission begins to increase.  Near point 3, the error signal abruptly changes sign and the cavity PZT changes direction indicating that the system has shifted from point 3 to point 4 as the growing plasma phase shift rapidly modifies the resonance detuning. As the plasma continues to grow, the cavity PZT moves to follow the shift in resonance, but the negative error signal suggests the loop cannot keep up.  Near the peak of transmission (1) the plasma phase shift saturates and the cavity PZT overshoots, leaving the cavity slightly off of resonance and causing the plasma to disappear.  The cycle begins anew with another rapid decrease in the cavity transmission.  To prevent this oscillation, the cavity can be locked slightly off of resonance (point 5).

These oscillation lineshapes are somewhat different that those seen in higher power systems due to the high finesse ($\mathcal{F}>2000$), which is responsible for the ringing in rising edge (and ringdown) of the cavity transmission.  Also this behavior is only observed for the through-nozzle gas geometry in the Ti:Sapphire system, which has a considerable interaction length with xenon gas and plasma.  We do not see these oscillations for the end-fire gas geometry at the intracavity pulse energy and gas pressures we can achieve with this system.  In our Yb-doped fiber laser system, we find that oscillations similar to those in \cite{Yost2011} arise for high gas pressures and average fsEC power above 1-2~kW.  In this case, we are able to lock slightly off resonance to reduce the plasma oscillations, as first observed in \cite{Allison2011}.  The increased XUV output under these conditions appears to come at the cost of increased amplitude noise of the fundamental in the cavity, which could have implications in the ultimate linewidth of the resulting XUV frequency combs \cite{Yost2009}.

To curtail the limitations caused by the HHG-generated plasma, one must minimize its interaction time with the optical driving field.  Use of a lower repetition rate oscillator gives the gas atoms and plasma more time to move through the interaction region, but this is highly undesirable for a frequency comb. In direct frequency comb spectroscopy, increasing the power per comb element increases the signal, and the increased comb spacing decreases the likelihood that multiple comb lines drive the same atomic or molecular transition simultaneously.  The remaining option is to decrease the transit time of atoms through the interaction region. The velocity of the mixture flowing from the nozzle is, $v_{gas} \propto \sqrt{T/m}$, where $T$ is the gas temperature and $m$ is the average gas mixture mass, thus indicating two practical methods to increase the velocity. The first is to increase the gas temperature, which can be achieved either through heating the gas at the source, or by heating the gas delivery nozzle.  In the latter case, we have observed that when a small quartz nozzle is placed near the intracavity focus, substantial heating of the nozzle occurs without catastrophic diffraction losses introduced to the fsEC.  However, it remains to be observed whether the additional nozzle heat can be transferred to the gas.  The second method is to use a gas mixture, with the second gas type having a much lower mass than the harmonic generating medium (such as a xenon/helium mixture).  The consequence of a mixture is that in order to maintain the same xenon density at the focus with a substantial increase in velocity, the gas flow must be considerably increased, potentially increasing the reabsorption of the generated harmonics and putting strain on the pumps for the vacuum system.

\section{Performance summary of fsEC XUV sources}
\label{sec:fsEC_results}

\begin{figure}[h!]
\centering
\resizebox{0.88\textwidth}{!}{%
\includegraphics{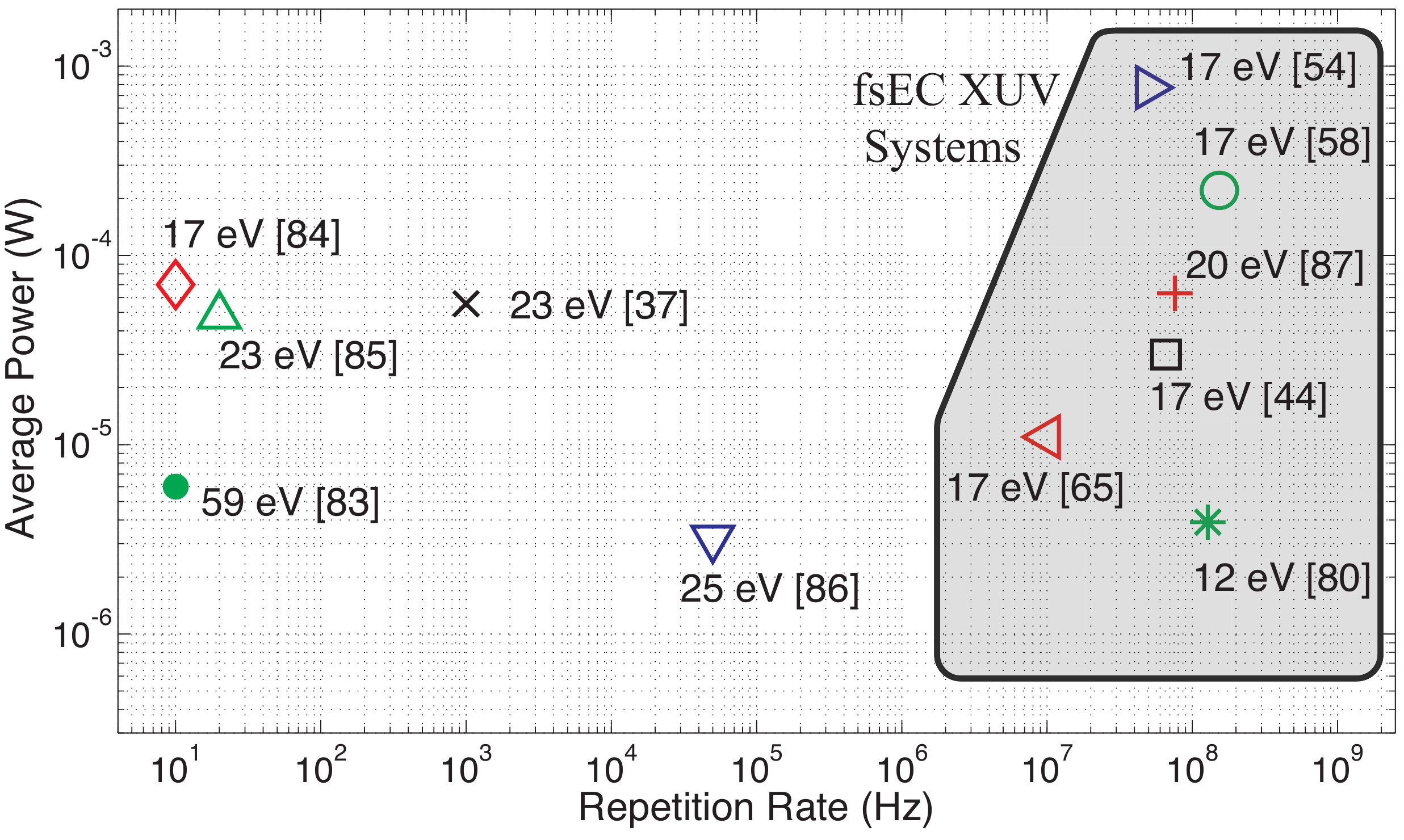}}
\caption{Comparison of performance of HHG XUV systems. Shown are the reported {\it generated} XUV powers before filtering of the fundamental or spectral separation of the XUV harmonics. An older version of this figure was first presented in \cite{Yost2011}.}
\nocite{Kim2008}
\nocite{Takahashi2002}
\nocite{Hergott2002}
\nocite{Constant1999}
\nocite{Hadrich2011}
\nocite{Ozawa2008}
\nocite{Lee2011}
\nocite{Yost2011}
\nocite{Hammond2011}
\nocite{Bernhardt2012}
\nocite{Lam2012}

\label{fig:HHGSummary}       
\end{figure}

In this section, we summarize some of the most recent developments in fsEC-based XUV sources.  To date, there are four classes of fsEC systems that have been used to generate XUV: 1) a high finesse fsEC at 800 nm seeded directly with a Ti:Sapphire mode-locked oscillator, 2) a lower finesse fsEC  at 800 nm seeded by an amplified Ti:Sapphire fs laser  3) an fsEC operating at 1~$\mu$m seeded by an amplified Yb-doped fiber laser system, and finally 4) a 500-nm fsEC seeded by a frequency doubled  Yb-doped fiber laser system.  A summary of these results along with other high repetition rate single-pass results (also using HHG) is shown in \fref{fig:HHGSummary}. Note that for this comparison all numbers were taken as the reported generated XUV power. For fsEC XUV systems this corresponds to the intracavity XUV power, while for traditional CPA, single-pass XUV systems \cite{Heyl2012} it is the power generated just after the gas target. In both approaches similar XUV losses are experienced when separating the XUV from the fundamental and/or spectrally resolving the XUV. A similar figure was presented in \cite{Yost2011}, which we have updated with the latest results.

\subsection{Ti:Sapphire Oscillator}
Of the four types of systems demonstrated, the first was a high finesse cavity seeded with a Ti:Sapphire mode-locked oscillator. In this first class of systems \cite{Jones2005,Gohle2005,Ozawa2008}, the high peak intensity requirements of HHG were met by using large enhancement factors made possible through careful management of cavity dispersion. Although some variation exists in these systems, depending on the particular configuration (including pulse energy, repetition rate, and the focusing geometry and enhancement factor achieved in the cavity) these types of systems typically generate harmonics with a plateau near the 11th or 13th harmonic of the 800 nm seed. Our version of such a system \cite{Hammond2011} utilizes a grating output coupler ($\lambda_g=200$ nm) for the harmonics, which has enabled us to measure $>$1 $\mu$W (spectrally resolved) in the 11th harmonic ($4\times10^{11}$ photons/sec), and we have observed the 13th harmonic, which is about a factor of five lower in power.  Given the simplicity of this approach, it is interesting to explore the uses of such a system.

\subsection{Amplified Ti:Sapphire}
An improvement over the single oscillator Ti:Sapphire source was reported \cite{Paul2008}, wherein the authors demonstrated a high- power femtosecond amplification cavity.  In this system, a 1 W output from a mode-locked oscillator is used to injection lock an actively pumped, Ti:Sapphire amplifier cavity to achieve an average power of 7~W while maintaining fs pulses at 50 MHz when pumped with 18.5 W at 532 nm.  The resulting pulse train can be compressed to its transform limit ($\sim70$ fs), and is subsequently used to seed a fsEC.  With an enhancement of 100-200 times, pulse energies as high as 25 $\mu$J are produced, with an intensity at the cavity focus in excess of $10^{14}$~W/cm$^2$.  XUV generation with this fsEC was recently reported \cite{Lee2011}, in which $\mu$W power levels are observed for harmonics as high as the 15th, and 77 ($\pm 15$) $\mu$W of light at 72 nm (11th harmonic) was out-coupled from the fsEC.  These impressive results are obtained at a significantly lower pulse energy ($\sim 15$ $\mu$J) than the maximum operating conditions because of limitations due to the intracavity plasma. This system utilizes a Brewster plate output coupler that is about 10\% efficient at this wavelength, which means that $\sim$0.8 mW of the 11th harmonic is generated intracavity.  This is the highest power reported to date from a fsEC-based XUV system.

\subsection{High Power Yb-doped Fiber Amplifier}
One of the most promising systems for extended-use, high flux fsEC-based XUV systems uses a seed from an amplified femtosecond Yb-doped fiber laser operating near 1~$\mu$m.  These systems offer favorable power scaling and pulse durations compatible with fsEC use.  The most advanced source of this type reported to date has a spectrum centered near 1070 nm and boasts 120 fs pulses, 80 W average power and 154 MHz repetition rate \cite{Yost2011}.  The 0.52~$\mu$J pulses from this system only need modest enhancements to achieve the level needed for HHG.  In fact the high power of this source exceeds the levels at which the fsEC exhibits nonlinear behavior and possible mirror damage.  Nevertheless, as high as 10 kW can be achieved in this system. Power scaling of a similar system has also been reported \cite{Pupeza2010a}, in which 18 kW was achieved using 200 fs pulses and as high as 72 kW was achieved with pulses of 2 ps duration. The system reported in \cite{Yost2011} uses an XUV grating output coupler ($\sim$10\% diffraction efficiency), and is capable of generating $>$200~$\mu$W while delivering $\sim$20 $\mu$W/harmonic.  At 20~$\mu$W of measured power, this is currently the highest power reported that is spectrally resolved and can be used directly in an experiment.

\subsection{Visible light fsEC via Frequency Doubled Yb-doped Fiber Amplifier}
A visible wavelength fsEC was recently reported \cite{Bernhardt2012}.  In this work, the authors frequency-double the output of a Yb-doped fiber laser system to generate light at 520 nm, which is then used to seed a fsEC. This approach takes advantage of the relatively efficient ($\sim$50\%) second harmonic generation (SHG) process to gain access to the dramatically improved HHG efficiency that results from the shorter fundamental wavelength, which has been demonstrated to have a wavelength scaling that varies as $\lambda^{-6.3\pm1.1}$ \cite{Shiner2009}. Of course, a shorter wavelength also causes a reduction in the cut-off harmonic generated in the HHG process, although this is not an issue for some applications requiring a VUV frequency comb.

This demonstration illustrates the significant fundamental and technical challenges to working with the shorter fundamental wavelength.  High reflectivity mirrors are not readily available at 520 nm, and while an enhancement of 200 was achieved for low input powers, the enhancement dropped to 75 for an input power of 4~W.  To even maintain this latter level of enhancement, the fsEC must be operated in an oxygen environment to prevent mirror degradation that occurs in vacuum, as was discussed in \sref{subsection:MirrorDamage}. This degradation phenomenon is similar in nature but dramatically more pronounced than that observed in 800-nm or 1040-nm based  fsEC systems. Due to VUV absorption, the oxygen environment of the cavity is a major problem for the harmonics generated, and in order to extract the harmonics from the cavity a Brewster plate is used as the fsEC output coupler that directs the harmonics into a small tube towards a differentially pumped detection chamber to minimize the optical path length in oxygen.  The system is reported to generate (out-couple) a power of 6.5 mW (3.1 $\mu$W) at 173 nm, 3.9$\mu$W (120 nW) at 104 nm, and 150 nW (5 nW) at 74 nm.  While the reported photon flux does not currently favor the SHG route to VUV generation (compared to using the same Yb-doped fiber source and an IR fsEC), this situation may improve greatly if the current limitations are better understood and resolved.

\section{Current and future fsEC applications}
\label{sec:fsEC_apps}
In this section we review current and future applications of VUV and XUV frequency combs produced with fsEC HHG generation. As the primary motivation for pursuing HHG in a fsEC is the generation of an XUV femtosecond frequency comb, the first studies reporting fsEC HHG sought to demonstrate the coherence of the harmonics. We begin with an overview of these experiments, after which we discuss the current and future applications of XUV frequency combs. Some of these applications require the coherence of the frequency comb, while others take advantage of the high repetition rate for low probability photoelectron events. Lastly, applications that benefit from the low-noise generation of below-threshold harmonics are discussed.

In the first two demonstrations of HHG in a fsEC, the coherence of the third harmonic was demonstrated.  The authors of \cite{Jones2005} observed a resolution bandwidth limited 1 Hz heterodyne beat frequency between the fsEC-generated 3rd harmonic of a Ti:Sapphire seed  and the third harmonic generated by conventional nonlinear optical crystals.  A second research group observed heterodyne beating between the fsEC third harmonic and the fifth harmonic generated via nonlinear optics from a Nd:YVO$_4$ laser at 1064 nm \cite{Gohle2005}.  The phase coherence between the Ti:Sapphire fsEC seed and the Nd:YVO$_4$ laser was established by phase-locking the second harmonic of the Ti:Sapphire laser to the third harmonic of the Nd:YVO$_4$.  These measurements were a very promising sign that a high degree of phase coherence is preserved in the HHG process even when generating several harmonic orders simultaneously through this highly nonlinear process.

More recently, a pulse-to-pulse coherence measurement was made on the 7th harmonic of a Yb-doped fiber fsEC system operating at 1070 nm \cite{Yost2009}.  In this VUV measurement, a cross-correlation interferometer was used with a one-pulse period delay, allowing the observation of interference between successive pulses from the fsEC source.  As a central part of this work, the authors investigate the role of multiple quantum pathways in the generation of below-threshold harmonics and their coherence properties, and they measure an interference that arises between different pathway contributions to the harmonic signal that vary with the fundamental intensity.  In essence, the intensity-dependent quantum path dependence for electrons participating in the HHG process represents an intensity-to-phase noise conversion that could impact the linewidth of the XUV frequency comb elements.  The authors conclude that even with a frequency comb source with 1\% amplitude fluctuations, high coherence XUV combs should be possible.  However, this intensity-to-phase noise conversion is expected to be worse for above-threshold harmonics, as the authors point out.

Very recently, the same group has used the 13th harmonic of their Yb-doped fiber laser system at 1070 nm to observe electric dipole (E1) transitions in argon at  $\lambda \sim82$nm, and the 17th harmonic to observe an E1 transition in neon, at $\lambda \sim63$nm \cite{Cingoz2012}.  An absolute frequency measurement of the argon transition establishes an uncertainty of this XUV absolute frequency measurement of $\pm 3$ MHz, limited by Doppler shifts that arise in the supersonic beam geometry of the neon and argon.  This measurement puts an upper bound on the linewidth of the 13th harmonic, and is still many orders of magnitude above what is believed to be the true frequency stability of the XUV comb, so work is currently underway to determine these limits.

Future measurements include dual comb XUV spectroscopy, in which two frequency combs with slightly detuned repetition rates are used to perform precision spectroscopy \cite{Schiller2002}.  The power of this technique is that when a heterodyne mixing of these two combs is performed, under the right conditions, the heterodyne beat between each pair of optical modes appears at a unique radio frequency, thus establishing a one-to-one mapping from optical frequencies to radio frequencies.  When this measurement is performed in a two-branch configuration, such as in a Mach-Zehnder interferometer, with an absorptive or dispersive sample in one branch, the amplitude and phase response of the sample can be determined across the optical frequency spectrum of the two frequency combs.  Extension of this technique to the XUV spectrum is now within reach \cite{Lee2011}, and should enable spectroscopy on a wide variety of new systems.  Furthermore, a practical limitation of XUV dual comb spectroscopy is that, in all likelihood, a second, independent fsEC XUV source will be required.  The dual comb measurements themselves, even without a spectroscopic sample will provide a comprehensive test of the XUV frequency comb to establish its ultimate frequency stability.

The high repetition rate of fsEC XUV sources is also a benefit for experiments such as ARPES, in which well-developed photoelectron statistics are required and low repetition rate systems require long acquisition times to perform.  As mentioned in the introduction, for an adequate XUV flux, the higher repetition rate translates to a lower XUV pulse energy that significantly reduces problems that may arise due to space charge effects on the photoelectron trajectories.  One challenge to the application of fsEC XUV sources to ARPES will be the spectral resolution of the source, where the spectral bandwidth of the entire frequency comb underneath a single harmonic is more important than that of the individual comb elements themselves. A recent demonstration of a fsEC seeded with 190 fs pulses at 1040 nm is expected to have a spectral width below $50$ meV in the XUV \cite{Lam2012} which is on the order of the spectral resolution most spectrometers used in ARPES experiments. Direct measurement of the harmonic spectral width by examining the Fermi edge of gold is now underway.

Yet another feature of the fsEC XUV frequency comb that will be useful in future experiments has nothing to do with its frequency comb structure.  XUV scattering experiments, such as nano-aerosol size metrology, will benefit from the low relative intensity noise (RIN) of the harmonics generated in fsEC.  Because the seed sources for fsEC's can have very low RIN, low intracavity RIN can also be achieved. With the help of high performance servo control loops for the cavity stabilization, the RMS noise can be limited to far less than 1\%, leading to nearly 1\% RMS amplitude noise on the harmonics generated in a fsEC with $\mathcal{F}>2000$ \cite{Hammond2011a}. These noise levels are expected to make such extinction measurements, first demonstrated on VUV radiation from a synchrotron \cite{Wilson2007}, much easier.

\section{Outlook}
In this review we have discussed the development of very high repetition rate ($>$ 50 MHz) XUV sources based on femtosecond enhancement cavities.  We covered the fundamental theory of their operation, details of their implementation, and some of the fundamental challenges to increasing the photon flux and running time of these sources. New fs oscillator developments also offer the possibility of simplifying existing experimental configurations for some applications.

High power Yb-doped fiber oscillators have been developed that could be used to replace multiple amplifier stages in current CPA systems.  Recently a dissipative soliton fiber laser delivering 140 nJ pulses at a repetition rate of 84 MHz and a nearly transform limited compressed pulse duration of 115 fs pulses was reported \cite{Lefrancois2010}.  This source produces pulses that are already sufficient as a seed source for a fsEC, and this compact arrangement is an attractive option to explore.  Given the power scalability of Yb-doped oscillator systems, it is also interesting to investigate the possibility of intra-oscillator-cavity HHG.  Indeed, one such proof-of-principle demonstration has been recently reported in a Ti:Sapphire oscillator cavity \cite{Seres2012}.

Increasing the running time of fsEC sources is largely made possible through the adherence to good-practice techniques for high-vacuum applications in order to minimize hydrocarbon contamination on the mirror surfaces.  This is somewhat challenging given the requirements that optical systems require regular adjustment, but vacuum-compatible opto-mechanics are becoming more readily available.  It is certainly the case that fsEC systems operating near 1 $\mu$m can be made to run with extended running times, but operation for extended periods with Ti:Sapphire systems near 800 nm, or visible light fsECs remains uncertain due to mirror/optic degradation issues.

Several challenges remain to further scaling of the generated XUV flux. With a fixed focal volume, further gains will not be achieved unless plasma can be flushed from the interaction region faster than is currently done.  This will decrease the fraction of neutral atoms that are ionized by the time they reach the center of the focal volume, where the peak intensity exists.  Further increases in XUV flux are possible by increasing the mode volume within the intracavity focus while holding the intensity constant.  The development of high power Yb-doped fiber amplifiers has made this possible for systems operating near 1 $\mu$m, and there is still some scalability available in current amplified Ti:Sapphire systems operating near 800~nm, but significant challenges still exist. It remains to be seen at what level plasma-related effects might limit further scaling of the output flux for larger mode volumes.  And finally, almost a full order of magnitude of further increase in the deliverable XUV flux can be gained through the development of more efficient output coupler designs.

A number of new experiments are currently underway that will demonstrate the capabilities of fsEC XUV sources. These include direct frequency comb spectroscopy, ARPES, XUV scattering measurements, and dual comb spectroscopy.  These applications utilize several of the key features of the fsEC source, including very high repetition rate, high frequency resolution, and the relatively low intensity noise of the XUV light produced in these sources.  These applications will benefit a variety of areas in science, and will provide critical tests of the sources themselves, such as establishing the ultimate linewidth achievable with the XUV frequency comb.

\ack 
We thank Rob Stead and Egor Chasovskikh for their work in the development of our fsECs, and R. Jason Jones and Jun Ye for useful discussions.  This research is supported by Natural Science and Engineering Research Council (NSERC), NRC-NSERC-BDC Nanotechnology Initiative, Canadian Foundation for Innovation, and British Columbia Knowledge Development Fund. 
\section*{References}
\bibliographystyle{ieeetr}
\bibliography{fsEC_Review.bib}

\end{document}